\documentclass[11pt,a4paper]{article}  
\usepackage{amsfonts, amsmath, amssymb, mathrsfs, balance, booktabs, caption, comment,  cmap,  environ, etoolbox, fancyhdr, float, fontaxes, geometry, graphics, hyperref, iftex, inconsolata, libertine, manyfoot, microtype, mmap, mweights,  nccfoots, refcount, setspace, textcase, totpages, trimspaces, upquote, url, xcolor, xkeyval, xstring}

\AtBeginDocument{%
  \providecommand\BibTeX{{%
    \normalfont B\kern-0.5em{\scshape i\kern-0.25em b}\kern-0.8em\TeX}}}

\usepackage[utf8]{inputenc}
\usepackage{graphicx}
\usepackage{amsmath,amsthm,amssymb}
\usepackage[english]{babel}
\usepackage{float}
\usepackage{subcaption}
\usepackage{fullpage}
    \usepackage{hyperref}
\usepackage{xcolor}
    \hypersetup{
        colorlinks,
        linkcolor={red!90!black},
        citecolor={green!80!black},
        urlcolor={blue!80!black}
    }

\usepackage{csquotes}
\usepackage[
    backend=biber,
    style=numeric,
    natbib=true,
    sortlocale=en_US,
    url=false, 
    doi=true,
    eprint=false
]{biblatex}
\usepackage{todonotes}

\usepackage[linesnumbered,ruled]{algorithm2e}
\usepackage{mathtools}
\usepackage{pgfplots}
\pgfplotsset{width=4.5cm,compat=1.9}
\usepackage{pgfplotstable}
\usepgfplotslibrary{groupplots}

\def\polylog{\operatorname{polylog}}
\DeclarePairedDelimiter\ceil{\lceil}{\rceil}

\newcommand{\blue}{\color{blue}}
\renewcommand{\O}{\mathcal{O}}
\newcommand{\C}{\mathscr{C}}

\pgfplotsset{
    every non boxed x axis/.style={} 
}

\title{On Practical Nearest Sub-Trajectory Queries under the Fréchet Distance}

\usepackage{authblk}
\author[1]{Joachim Gudmundsson}
\author[2]{John Pfeifer}
\author[3]{Martin P. Seybold}
\affil[1]{
     \small{University of Sydney, School of Computer Science, Sydney,
Australia.}
     \footnotesize{joachim.gudmundsson@sydney.edu.au}
}
\affil[2]{
     \small{University of Sydney, School of Computer Science, Sydney,
Australia.}
     \footnotesize{johnapfeifer@yahoo.com}
}
\affil[3]{
     \small{University of Vienna, Faculty of Computer Science, Vienna,
Austria.}
     \footnotesize{martin.seybold@univie.ac.at}
}
\date{Published: \href{https://dl.acm.org/doi/10.1145/3587426}{02 May 2023}}

\begin{document}

\maketitle
\begin{abstract} 
We study the problem of sub-trajectory nearest-neighbor queries on polygonal curves under the continuous Fréchet distance.
Given an $n$ vertex trajectory $P$ and an $m$ vertex query trajectory $Q$, we seek to \emph{report} a vertex-aligned sub-trajectory $P'$ of $P$ that is closest to $Q$, i.e. $P'$ must start and end on contiguous vertices of $P$.
Since in real data $P$ typically contains a \emph{very large} number of vertices, we focus on answering queries, without restrictions on $P$ or $Q$, using only precomputed structures of ${\mathcal{O}}(n)$ size.

We use three baseline algorithms from straightforward extensions of known work, however they have impractical performance on realistic inputs.
Therefore, we propose a new Hierarchical Simplification Tree data structure and an adaptive clustering based query algorithm that efficiently explores relevant parts of $P$.
The core of our query methods is a novel greedy-backtracking algorithm that solves the Fréchet decision problem using $\O(n+m)$ space and $\O(nm)$ time in the worst case.

Experiments on real and synthetic data show that our heuristic effectively prunes the search space and greatly reduces computations compared to baseline approaches.
\end{abstract}


\section{Introduction} \label{sec:intro}
Increasingly sophisticated and inexpensive movement capture devices have led to exponential growth in trajectory data over the past decade.
Large quantities of trajectory data now exist in diverse fields of study such as sports analysis, human body movement, and vehicle tracking.
This has necessitated the need for algorithms that answer trajectory-based queries efficiently.
These underlying drivers, along with difficulties in constructing methods that produce exact results, have led to trajectory algorithms~\cite{de2013fast,unpublished1} that focus on reducing computational complexity by employing \emph{approximation} algorithms.
However, these methods may not be sufficient in settings where meaningful investigation requires exact results.

In this work, we study the problem of \emph{exact} sub-trajectory nearest-neighbor searches.
Given a $d$ dimensional input trajectory $P$ of size $n$, and query trajectory $Q$ of size $m$, with $m \ll n$, the problem is to find a contiguous sub-trajectory $P'$ within $P$ that is closest to $Q$ under the continuous Fréchet distance~\cite{alt95}.
$P'$ must start and end on vertices in $P$, and $P'$ is inclusion minimal, i.e., it cannot be shortened and still be closest to $Q$.

The problem of sub-trajectory proximity searches under the continuous Fréchet distance has been studied from a theoretical point of view.
In particular, work exists on approximation algorithms for the range counting problem~\cite{de2013fast,unpublished1} and a data mining problem~\cite{buchin2011detecting} where one searches for similarly close sub-trajectories within input $P$.
However, the exact search problem is less studied. 

We study exact algorithms and turn our attention towards pragmatic methods that prune the search space and reduce computations.
Our goal is a \emph{practical} nearest-neighbor search algorithm which gives results that are correct and exact.
In our setting, the solution must: (i) return a nearest-neighbor with no restrictions on $P$ or $Q$, (ii) have low preprocessing time, and (iii) use linear storage.

\subsection{Contribution and Paper Outline} \label{ssec:cont}


We present three different baseline algorithms (cf. Section~\ref{sec:base}) that return exact or approximate vertex-aligned sub-trajectory results, obtained by slight modifications of existing work to our problem setting.
These are 
(B1) adjusting the decision algorithm of Alt and Godau~\cite{alt95}, 
(B2) adapting the approximate Fréchet distance algorithm of Driemel et al.~\cite{driemel2012approximating}, and 
(B3) using the metric indexing technique from~\cite{gudmundsson2021practical}.
These baseline algorithms have fast query times for certain types of inputs and queries, however all have drawbacks, such as impractically large data structures and query search spaces, on realistic input (cf. Figure~\ref{fig:realistic_inputs}).

We counter those drawbacks with a new data structure and sub-trajectory query algorithms (cf. Section~\ref{sec:new}) that are practically more efficient in searching the input space on real and synthetic data sets.
The method uses three techniques, with the third expanding on the first two and also offering the best experimental performance.

Our first technique (cf. Section~\ref{ssec:algo1}) is a novel greedy method, with backtracking, that searches the freespace diagram of $P$ and $Q$ to solve the sub-trajectory Fréchet decision problem in $\O(n+m)$ space and $\O(nm)$ pointer machine operations in the worst case.
Though improving on B1 query times, the technique alone does not scale substantially better for very large input trajectories. 

To enable metric pruning techniques on the sub-trajectories of $P$, we introduce the Hierarchical Simplification Tree (HST) to quickly derive relevant sub-trajectory clusters in the query phase (cf. Section~\ref{ssec:hst}).
The HST stores simplifications 
of the input trajectory $P$ at various resolutions. 
Its space is $\mathcal{O}(n)$ and construction time is $\mathcal{O}(n D)$, where $D \leq \min \left\{ n, \O \left( \log \Delta(P) \right) \right \} $ and spread $\Delta(P)$ is the ratio between the largest and smallest Euclidean distance in the set of vertices from $P$.
Though HST sub-trajectory clusters lack the strong quality guarantees of B3, the construction time is four to five \emph{orders of magnitude} faster in practice.

Our second technique (cf. Section~\ref{ssec:algo2}) performs $\O(\log \Delta (P))$ iterations of a breadth-first search that derives and prunes clusters of sub-trajectories from the HST using the triangle inequality.
Though irrelevant clusters of sub-trajectories are pruned based on distance computations that only involve simplified sub-trajectories of $P$, the technique alone improves only sporadically over B1 and B2.

Our integrated, third technique combines both and additionally uses simplifications to accelerate the freespace technique and heuristics~\cite{bringmann2019walking,gudmundsson2021practical} to accelerate the pruning technique (cf. Section~\ref{ssec:algo3}).

Experiments show that the method is suited to handle very large inputs on standard laptop hardware, has orders of magnitude faster construction times, and query times that improve on all baseline methods.
This includes B3, the recent practical 
metric index method of \cite{gudmundsson2021practical} that is specifically designed for trajectory proximity searches, but results in a data structure of quadratic size in the sub-trajectory setting (cf. Section~\ref{sec:exp}).

\section{Related Work} \label{sec:rel_work}
The sub-trajectory \emph{nearest-neighbor} search problem can be solved in $\mathcal{O}(mn \log mn)$ time, whereas the sub-trajectory \emph{range} search can be computed in $\mathcal{O}(mn)$ time, both based on simple modifications to Alt and Godau's~\cite{alt95} decision algorithm.
Unlike this decision algorithm, recent versions with improved time bounds or improved practical behavior do not allow clear extensions to the sub-trajectory decision problem.
To our knowledge, there is also no available implementation of the (mildly) sub-quadratic decider~\cite{buc217}, whose space bound matches its time bound on the pointer machine.
The practical, recursive decider in~\cite{bringmann2019walking} supports the basic decision problem and can be executed in linear space.
However, extensions for the sub-trajectory setting are unclear and the time bound of the recursive method is super-quadratic (i.e. checking `simplicity' of a box boundary~\cite[Algorithm~$2$]{bringmann2019walking} does not have an $\O(1)$ time bound).

Two studies~\cite{de2013fast,unpublished1} give approximation algorithms for the \emph{counting} version of the sub-trajectory range problem under the continuous Fréchet distance.
De Berg et al.~\cite{de2013fast} describe a multi-level partition tree that takes $\mathcal{O}(n^3 \log n)$ time to construct and uses $\mathcal{O}(s \polylog n)$ space, where $n \le s \le n^2$.
The data structure can only handle single segment queries in $2$D that must be longer than $6\tau$, where $\tau$ is the range input parameter.
The query time is $\mathcal{O}(n/\sqrt{s} \polylog n)$, and it counts all sub-trajectories up to distance $\tau$ from $Q$, but the reported value may also contain those up to $(2+3\sqrt{2})\tau$ distance from $Q$.
Gudmundsson and Tridgell~\cite{unpublished1} present two algorithms that improve on~\cite{de2013fast}.
The first algorithm computes a $2\tau$-maximally simplified curve from $P$ in quadratic time, and the query algorithm walks along the simplified curve to produce a result in $\mathcal{O}(n)$ time.
$Q$ is restricted to have a constant complexity and each segment must have length of at least $8\tau$, and the count may include sub-trajectories up to distance $3\tau$ from $Q$.
The second algorithm pre-computes a multi-level data structure in $\mathcal{O}(n^3 \log n)$ time with $\mathcal{O}(n \polylog n)$ space.
Its query algorithm works for $d$ dimensional trajectories and takes $\mathcal{O}(n^{1-1/d} \polylog n)$ with an error of $3\sqrt{d}\tau$.
Both studies above bound query times, at a cost of quadratic or larger pre-processing time, limitations to $Q$, and approximate results.
They also provide counts and do not report sub-trajectory start/end points. 

In~\cite{buchin2011detecting}, Buchin et al. study a data mining problem that detects similar $2$D sub-trajectories in $P$ under the discrete and continuous Fréchet distance measures.
One result is an optimization algorithm that finds the \emph{maximum} sub-trajectory length $\bar{l}$ where there are at least $\bar{c}$ sub-trajectories and the continuous Fréchet distances between them are at most $\tau$. 
The algorithm gives a $2$-distance approximation and runs in $\mathcal{O}(n^2 \bar{l})$ time and $\mathcal{O}(n \bar{l}^2)$ space.

Driemel and Har-Peled~\cite{driemel2013jaywalking} describe a linear size data structure for $P$, that takes a \emph{single} segment $Q$ and point indices $x$ and $y$ as input, and returns a $(1+\varepsilon)$-approximate Fréchet distance between $Q$ and the sub-trajectory $\langle p_x, \ldots , p_y\rangle$, in $\mathcal{O}(\log n \log \log n)$ time. 
There is also work~\cite{ber17,driemel2017locality} on approximate nearest-neighbor searches under the discrete and continuous Fréchet distance measures that find the closest trajectory within an input set, however the methods do not search for sub-trajectories and have exponential data structure size.
See~\cite{rangeLowerBound} for lower bounds for approximate range searching.

The recent work~\cite{gudmundsson2021practical} contributes a practical approach for \emph{exact} proximity searches on sets of input trajectories based on clustering with strong quality guarantees and query algorithms that exploit potentially low `intrinsic dimensionality'~\cite{kargerR02, GuptaKL03} of the data sets for metric pruning.
However, the method does not extend well to our problem since the data structure size is quadratic in the sub-trajectory setting.

\section{Preliminaries} \label{sec:prelim}

We now provide definitions for trajectories, the continuous Fréchet distance, the nearest-neighbor sub-trajectory search problem, and trajectory simplifications. 

\subsection{Trajectories} \label{ssec:traj}
An input trajectory $P$ of size $n$ is a polygonal curve through a contiguous sequence of $n$ vertices $\langle p_1, \ldots , p_n\rangle$ in $\mathbb{R}^d$, where each vertex pair $p_i, p_{i+1}$ is connected by a straight-line segment $\overline{p_i p_{i+1}}$.
The \emph{length} of $P$ is the sum of the Euclidean lengths of all its segments. 
A query is a trajectory $Q$ of size $m$, $Q = \langle q_1, \ldots , q_m\rangle$.
%
%
A sub-trajectory of $P$ is denoted $P'$, and is \emph{vertex aligned}, meaning its first and last vertices are vertices of $P$, i.e. $P' = \langle p_i, \ldots , p_j\rangle$ with $1 \le i \le j \le n$.

\subsection{Continuous Fréchet (CF) Distance} \label{ssec:frechet_dist}
The continuous Fréchet distance between two trajectories $P$ and $Q$ can be envisaged as the minimum `leash length' required between a person walking monotonously along $P$, and their dog walking monotonously along $Q$.
We associate $P$ with its natural parameterization ${P : [0,1] \to \mathbb{R}^d}$, which maps positions relative to $P$'s length to spatial points -- e.g. $P(0.5)$ is the mid-point of $P$.
A continuous, monotonous map $f : [0,1] \to [0,1]$ is called a reparameterization, if $f(0) = 0$ and $f(1) = 1$,
with $\mathcal{F}$ representing the set of all reparameterizations.
The continuous Fr\'echet distance is defined as
\begin{equation}
CF(P,Q) = \inf_{f,g \in \mathcal{F}} \max_{\beta \in [0,1]}
\Big\lVert
P \Big(f\left(\beta\right) \Big) - Q \Big(g(\beta) \Big) 
\Big\rVert, 
\end{equation}
where $\lVert \cdot \rVert$ is the Euclidean norm in $\mathbb{R}^d$.
We refer to the continuous Fr\'echet distance as $CF$ or \emph{distance} throughout this work, when it is contextually clear.
$CF$ can be computed in $\mathcal{O}(mn \log mn)$ time using the algorithm of~\cite{alt95},
which performs multiple calls to an $\mathcal{O}(mn)$ time decision procedure, denoted $DP(P,Q,\tau)$, that tests if $CF$ is at most $\tau$.

The continuous Fréchet distance is a (pseudo) metric and hence can be used in metric indexing schemes~\cite{beygelzimer2006cover,gudmundsson2021practical}. 

\subsection{The Nearest Sub-Trajectory Problem} \label{ssec:problem_def}
Given an input trajectory $P$ and query trajectory $Q$, with $m \ll n$, the problem is to find a sub-trajectory $P'$ that is closest to $Q$ under the continuous Fréchet distance.
Both the sub-trajectory $P'$ and the CF distance between $P'$ and $Q$ must be reported.

In our setting, there can be more than one sub-trajectory of $P$ that is closest to $Q$.
In this case, we report the inclusion minimal~\cite{de2013fast} sub-trajectory, i.e., $P'$ cannot be \emph{shortened} and still be closest to $Q$.
If there is more than one minimal inclusion result, then report one of them.

We also define the concept of inclusion maximal $P'$, which is used in our third query algorithm (cf. Section~\ref{ssec:algo3}).
In this case $P'$ cannot be \emph{lengthened} in size and still be closest to $Q$.

\begin{figure*}
    \centering
    \includegraphics[width=\columnwidth]{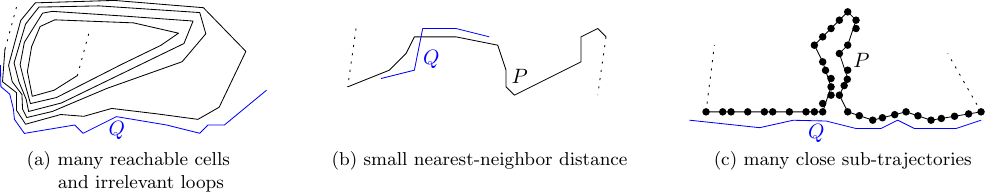}
    \caption{Examples of realistic inputs that lead to high query times in baseline methods $1$, $2$, and $3$.}
    \label{fig:realistic_inputs}
\end{figure*}

\subsection{Greedy Ball Simplification} \label{ssec:simplifications}
We use the trajectory simplification algorithm by Driemel et al.~\cite{driemel2012approximating}, which takes $P$ and a simplification error $\varepsilon$ as input and returns a simplified trajectory.
It is denoted $s(P,\varepsilon)$, and works as follows.
Record the initial vertex $p_1$, and set it to the current vertex.
Scan the next vertices, in order, until the first $p_i$ that is further than $\varepsilon$ away from the current vertex.
Record $p_i$ and set it to the current vertex.
Repeat until reaching $p_n$, and record $p_n$.
The recorded vertices are the simplification result of $P$.
The algorithm runs in $\O(n)$ time, and the simplification result has $CF(P,s(P,\varepsilon))\leq \varepsilon$.

The greedy ball algorithm has useful properties: (i) it snaps all contiguous vertices to $p_i$, if they are within $\varepsilon$ distance from $p_i$, (ii) the simplified curve always contains the first and last vertices of $P$, and (iii) all simplified segments (except the last one) have length greater than $\varepsilon$.
Such \emph{local simplifications} are particularly useful in obtaining a Hierarchical Simplification Tree~(Section~\ref{ssec:hst}), which allows pruning of sub-trajectory clusters during the search~(Section~\ref{ssec:algo2}).

\section{Baseline Query Algorithms} \label{sec:base}
There are known algorithms for computing the continuous Fréchet distance exactly~\cite{alt95} or approximately~\cite{driemel2012approximating,driemel2013jaywalking} and methods for computing exact~\cite{gudmundsson2021practical} or approximate~\cite{ber17,driemel2017locality} nearest-neighbor from an input trajectory \emph{set}. 
With relatively small and straightforward modifications, some of these algorithms can be modified to solve the problem we study, the vertex aligned sub-trajectory nearest-neighbor search.
We discuss three \emph{baseline} algorithms
and then provide realistic examples where 
they
are ineffective.
The three baseline algorithms broadly cover known methods, and are the starting point for the design of our novel data structure and algorithm that overcomes baseline performance issues (see Figure~\ref{fig:realistic_inputs}).

\subsection{Baseline 1 - Freespace Decider} \label{ssec:base1}
A relatively straightforward modification to Alt and Godau's decision procedure~\cite{alt95} yields the first baseline sub-trajectory nearest-neighbor algorithm.
The procedure decides if $P$ and $Q$ have at most a distance of $\tau$: 
$DP(P,Q,\tau)$ returns $true$ if $CF(P,Q) \le \tau$, otherwise it returns $false$.
We first provide a short description of the classic decision procedure algorithm, then explain the modification.

The $DP$ algorithm computes a \emph{freespace diagram}~\cite{alt95}, which is a grid that shows all pairs of points on $P$ and $Q$ that are at most $\tau$ distance apart:
$$ FS(P,Q) = \Big\{ (s,t) \in [1,m] \times [1,n] \: \Big| \: \lVert q_s - p_t \rVert \le \tau \Big\},$$
where $s$ and $t$ are positions (on vertices or segment interiors).
$FS$ is discretized with $m$ vertical grid lines corresponding to $Q$ vertices, and $n$ horizontal grid lines corresponding to $P$ vertices, with $(q_1,p_1)$ at the bottom-left grid corner, and $(q_m,p_n)$ at the top-right grid corner.
There are $(m-1) \times (n-1)$ grid cells, each representing two segments, one from $P$ and one from $Q$,
and~\cite{alt95} shows that the freespace for a given cell is computed by determining the intersection of an ellipse and the cell boundary, i.e., in constant time.

A \emph{reachable} point ($q_s,p_t$) in the $FS$ is defined as a point that has a monotone path, through freespace, from ($q_1,p_1$) to ($q_s,p_t$), where points on $P$ and $Q$ continuously increase along the path (e.g., they cannot `walk backwards').
Alt and Godau show that if a monotone path can be constructed through reachable space from ($p_1,q_1$) to ($q_m,p_n$), then $DP(P,Q,\tau) = true$, otherwise it is $false$.

An algorithm for computing the reachable space is as follows.
$FS$ grid cells are scanned row-by-row starting at the bottom row, and within a row cells are searched from left-to-right.
Reachable space is propagated from ($q_1,p_1$) along a monotone path, as each cell's freespace is computed.
If for a given grid row, there is no reachable space along the top boundary, then stop and return $false$, else if ($q_m,p_n$) is reached return $true$.
The algorithm has $\O(mn)$ runtime and can be implemented in $\O(\min(m,n))$ space.

The classic decision procedure algorithm above can be modified to answer the following question: decide if \emph{any sub-trajectory} $P'$ within $P$ has at most Fréchet distance $\tau$ from query $Q$: $DP_{FD}(P,Q,\tau)$ returns $true$, if for any $P'$, $CF(P',Q) \le \tau$, otherwise it returns $false$.

\begin{figure}\centering
    \includegraphics[width=\textwidth]{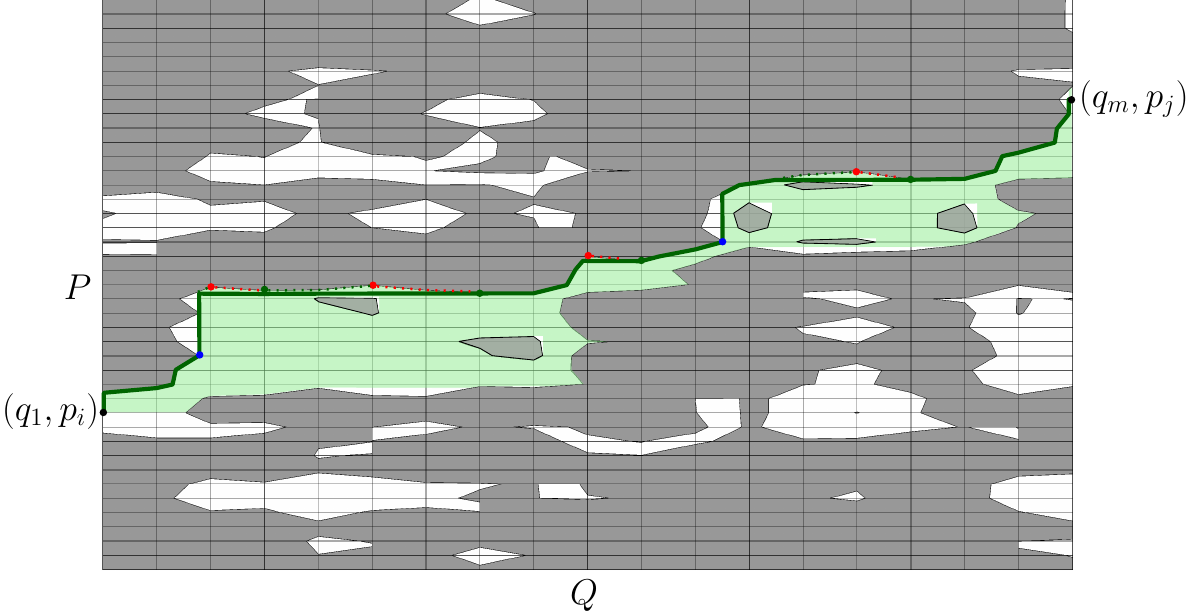} \\
	\caption{Freespace diagram example showing a trajectory $P$ ($|P|=41$) and query $Q$ ($|Q|=19$) from the Pigeon~\cite{pigeon16} data set. 
	In this example, point $(q_m,p_j)$ is reachable from point $(q_1,p_i)$ via the canonical path (green).
	Also shown are stopper points (red), trigger points (green), emission points (blue), and the reachable space in green (see Section~\ref{ssec:algo1}).
	Note that for simplicity, freespace cell plots show straight edges (technically they are ellipses).
	}
	\label{fig:fs2}
\end{figure}

Intuitively, one just needs to modify the algorithm above to search for a monotone path from \emph{any} freespace starting on the left side of the freespace diagram ($q_1$) to \emph{any} reachable space ending on the right side of the freespace diagram ($q_m$).
If such a path exists, then it follows there is a $P'$ (that starts on $p_i$ and ends on $p_j$) such that $CF(P',Q) \le \tau$. 
For example the green path in Figure~\ref{fig:fs2}.

We modify the classic decision procedure algorithm as follows.
Scan the matrix cells column-by-column starting at the left-most column, and within a column cells are searched from bottom-to-top.
Propagate reachable space the same as above.
If for a given column, there is no reachable space on the right boundary of the column, then stop and return $false$, else if reachable space is propagated to any point on the right side of the freespace diagram ($q_m$), then return $true$.
Note that the search is done column-by-column. 
The decision algorithm has $\mathcal{O}(mn)$ runtime and can be implemented in $\mathcal{O}( n )$ space.

In our setting $P'$ is vertex aligned, so vertices $p_i$ and $p_j$ must be in the reachable space, which must be checked in every call to $DP_{FD}$.
With the $DP_{FD}$ algorithm above it is trivial to check if one or more \emph{end} points in $P'$ is in reachable space.
To check if one or more \emph{start} points in $P'$ is in reachable space, search the reachable space in reverse.

Similar to the continuous Fréchet distance computation~\cite{alt95}, the sub-trajectory nearest-neighbor algorithm, denoted $CF_{FD}(P,Q)$, calls $DP_{FD}(P,Q,\tau)$ multiple times on a set of critical values, and can run in $\mathcal{O}(mn \log mn)$ time.
The distance and minimal inclusion sub-trajectory are returned for the case where $DP_{FD}(P,Q,\tau) = true$ and $\tau$ is minimal.

The Baseline $1$ algorithm 
has two primary issues in our setting, both of which are a consequence of the large size of $P$.
The first issue is that the parametric search uses $\Theta(mn)$ space which is unfortunate since in our setting $n$ can be very large.
A pragmatic way to avoid this is using a binary search on numeric digits of the distance value until a user-defined precision is reached.
The second issue is that the algorithm visits \emph{all reachable cells}, which can be numerous if $P$ has high sampling density (see Figure~\ref{fig:realistic_inputs}~(a)).

\subsection{Baseline 2 - Decider on Simplified Curves} \label{ssec:base2}
Inspired by the algorithm of Driemel et al.~\cite{driemel2012approximating}, we describe a simple method that reports a $(1+\varepsilon)$-\emph{approximate} nearest sub-trajectory.
The basic idea is to start with a large simplification error $\gamma$ on $P$ and reduce the error until an approximate result is found.
Instead of the binary search on the distance values from a well-separated pair decomposition of the points of $P$ and $Q$ (see Section~$3.3.3$ in \cite{driemel2012approximating}), we use a simple \emph{exponential search} over the range of possible values for the simplification threshold.

Start with iteration $i=1$ and set $\gamma = \max(reach(P),reach(Q))$, where $reach\big(\langle p_1, \ldots, p_n\rangle \big) = \max_j  \lVert p_j - p_1\rVert$ denotes the maximum Euclidean distance between the start and the other trajectory points~\cite{ber17}.
Simplify $P$ to obtain $P_\gamma = s(P,\gamma)$.
Compute the nearest-neighbor distance $\alpha = CF_{FD}(P_\gamma,Q)$, using the Baseline~$1$ algorithm.
If $(\alpha + \gamma) / (\alpha - \gamma) \le (1+\varepsilon)$ then stop and return the (inclusion minimal) sub-trajectory from $CF_{FD}$.
Otherwise set $\gamma$ to $\gamma/2^i$, increment $i$, and repeat on this resolution.

Note that the search for $\alpha$ is particularly demanding for values close to $\gamma(1+2/\varepsilon)=:\tau$.
Hence, we use the following criteria to avoid unnecessary precise estimation of $\alpha$ in the $CF_{FD}$ algorithm.
A call $DP_{FD}(P_\gamma,Q, \cdot)$ generates an upper or lower bound on $\alpha$, as discussed in Section~\ref{ssec:base1}. 
If $LB(\alpha) \geq \tau$, then stop and return the distance $LB(\alpha)$.
If $UB(\alpha) < \tau$ or $[LB(\alpha),UB(\alpha)] \subseteq [\tau/2,2\tau]$, then stop and proceed to the next finer resolution.
The value of $2$ in the latter criteria is a heuristic choice; the optimal value depends on the time trade-off between a $DP_{FD}$ call on resolution $\gamma$ and one on resolution $\gamma/2^i$.

The approximation algorithm, denoted $CF_{SC_\varepsilon}(P,Q)$, runs faster than the Baseline~$1$ algorithm if there are many sub-trajectories of $P$ that are close.
On queries with \emph{very small} nearest neighbor distances however, the approximation algorithm incurs substantial overhead compared to Baseline~$1$ (see Figure~\ref{fig:realistic_inputs}~(b) and Section~\ref{ssec:exp_results}).

\subsection{Baseline 3 - CCT Metric Index} \label{ssec:base3}
Since the Fréchet distance is a (pseudo) metric, one can apply metric indexes for searching the nearest neighbor in the set of sub-trajectories.
Metric indexes typically cluster inputs via a metric ball or bisector plane, and examples include the M-Tree~\cite{ciaccia1997} which reduces disk I/O accesses, or the Cover Tree~\cite{beygelzimer2006cover} whose nearest-neighbor search is bounded in terms of the expansion constant~\cite{kargerR02}.

Fortunately, one can answer sub-trajectory nearest-neighbor queries
efficiently
with the Cluster Center Tree (CCT)~\cite{gudmundsson2021practical}, a structure that is specifically designed to cluster and search on trajectories under the CF distance.
In our setting, take all $(^{n}_{2})$ pairwise sub-trajectories $P' \in P$ and insert them into the CCT.
Then, simply execute the CCT $kNN$ query algorithm for query $Q$ and $k=1$.
The query result will contain the closest vertex aligned sub-trajectory of $P$ to $Q$.

The CCT can provide a more favorable clustering on the set of sub-trajectories compared to our proposed HST, however it has two main drawbacks in our study setting.
This first is that the CCT input set contains $\binom{n}{2}$ trajectories, which results in a CCT data structure size of $\Theta(n^2)$.
This limitation means that in practical settings one must have input trajectories of small sizes (e.g., $|P| < 5000$).
Moreover, the CCT construction algorithm may require many $CF$ calls which impacts the pre-processing time.
The second drawback is when $P$ has high intrinsic dimensionality, i.e., there are many sub-trajectories in $P$ that are close to $Q$ (see Figure~\ref{fig:realistic_inputs}~(c)).
In this case the CCT search algorithm's pruning is less effective which results in more CF distance computations.

\section{Proposed Query Algorithms} \label{sec:new}
This section describes three algorithms for computing the nearest sub-trajectory.
The first is a greedy algorithm $DP_{GD}(P,Q,\tau)$ for deciding if $P$ contains a sub-trajectory with a CF distance of at most $\tau$ (cf. Baseline~$1$).
The second algorithm uses the new HST data structure and a breadth-first-search, which extends methods from metric indexing (e.g. the CoverTree~\cite{beygelzimer2006cover}) to clusters of sub-trajectories.
The third algorithm combines both, which results in a new method that addresses issues highlighted in the Baseline algorithms.

\subsection{Algorithm 1 - Greedy Decider} \label{ssec:algo1}
There are methods to obtain answers to the decision problem that are practically faster than the aforementioned Dynamic Program of Alt and Godau~\cite{alt95}.
Some are based on linear time heuristics that only fall back to the Dynamic Program if the heuristic is inconclusive~\cite{bal17,buch17,dut17}.
Beside those, the work of Bringmann et al.~\cite{bringmann2019walking} uses a divide-and-conquer approach to compute reachable sections of the freespace, with pruning rules that stop recursions early if a sub-matrix boundary is entirely reachable from the lower left corner or the lower left corner is separated from the upper right corner.
However, 
it is quite unclear if one can modify this approach for the sub-trajectory decision problem.

Our proposed greedy method uses backtracking to solve the decision problem exactly, requiring only pointer machine operations when working on the two lists of trajectory points.
We first describe the method for deciding if there is a monotone path from the lower left corner $(1,1)$ to the upper right corner $(m,n)$, and then discuss the modification required for deciding sub-trajectories.

For intuition, imagine there is a metal ball that starts in $(1,1)$ and that there is a strong magnet above and weaker magnet to the right of the freespace diagram.
The magnets attract the metal ball which moves in the freespace and primarily follows the \emph{boundary} between free and non-freespace.
The metal ball's reachable monotone path is tracked as it attempts to reach $(m,n)$.
 
We define the \emph{canonical path} to a (monotonously) reachable point backwards as the path that always chooses the highest reachable predecessor point to reach the last chosen point.
The canonical path to $(m,n)$ consists only of sections that are: (i) vertical, (ii) follow a boundary that is non-free above the path, or (iii) horizontal.
Note that every end of a horizontal passage coincides with a point on the freespace boundary.
For example, the green path in Figure~\ref{fig:fs2} shows the canonical path from $(q_1,p_i)$ to $(q_m,p_j)$  and all points that are monotonously reachable from $(q_1,p_i)$ in green.

Our decider searches for the canonical path to $(m,n)$ with a sweep over the freespace, i.e. we successively replace some suffix of a canonical path to obtain the canonical path to the next point.

\begin{figure}[h]
    \centering
    \includegraphics[ width=\textwidth ]{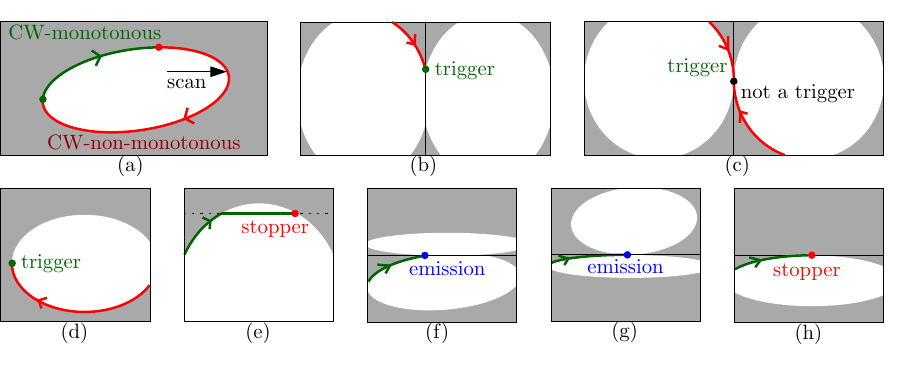}
    \vspace{-1.2cm}
    \caption{Clockwise monotone and non-monotone boundary traversal~(a), and examples for the definitions of trigger~(b,~c,~d), stopper in the presence of a ceiling~(e), stopper in the absence of a ceiling~(h), and emission points (f,~g).}
    \label{fig:mon-bdry}
\end{figure}

The clockwise (CW) traversal of a boundary curve (between free and non-free space) in any cell partitions its boundary in sections where the traversal is monotonous and non-monotonous (see Figure~\ref{fig:mon-bdry}~(a)).
We use the term \emph{stopper} for points on the boundary where the traversal changes from monotonous to non-monotonous and \emph{trigger} for points that switch from non-monotonous to monotonous (cf. Figure~\ref{fig:mon-bdry}).
Points on the top interface of a cell that are also on a monotonous section of the freespace boundary are called an \emph{emission} point if the space above is free.

Next, we describe the algorithm.

\subsubsection{Searching from $(1,1)$ to $(m,n)$}
The algorithm records, in two alternating states, the movement of the ball inside the freespace.
State $1$ is monotone movement, according to precedence of the magnets, and State $2$ is non-monotone movement along the boundary curve (with magnets `turned off').
Movement during State~$1$ is vertically through freespace, along a section of the boundary curve, or
along a horizontal ceiling line (e.g. $y=n$).
We record the canonical path by appending to a doubly connected list.
The search also uses a stack, storing trigger points, that initially contains only the target point $(m,n)$.
We call the $y$-coordinate of the peak element of the stack the (active) \emph{ceiling} and we fix the \emph{floor}, i.e. $y=1$.

The algorithm starts in State $1$, from point $(1,1)$, and moves the ball greedily upwards (using vertical ray shooting) and then monotonously along a boundary or ceiling.
If an emission point is reached on the boundary, then the ball moves vertically until it reaches a boundary or the ceiling.
This continues until either the target $(m,n)$ or a stopper is reached (see Figure~\ref{fig:mon-bdry} (e) and (h)). 

\begin{figure}[p]
    \centering
    \includegraphics[scale=1.0]{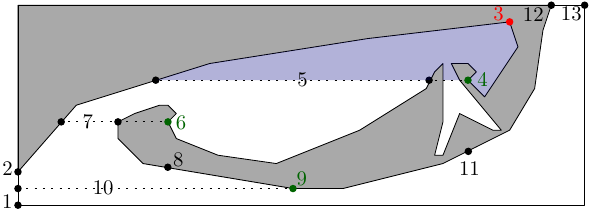}
    \caption{Greedy decider traversal example.
    {\normalfont
    Begin in state 1. Start at point 1, go up to 2, then to stopper 3. Switch to state 2. Go to trigger 4, then shoot ray 5 which hits a boundary, so push trigger 4 onto the stack. Go to trigger 6, then shoot ray 7 which hits a boundary, so push trigger 6 onto stack. Go to 8, pop trigger 6 off stack. Go to trigger 9, shoot ray 10 which reaches 9. Switch to state 1. Go to 11, pop trigger 4 off stack. Go to 12, and 13, and return $true$.
    The shaded blue area shows freespace that is pruned when trigger $4$ is encountered. 
    }
    }
    \label{fig:ex_traversal}
\end{figure}

If the monotonous path ends at a stopper $(x,y)$, then switch to State $2$ and follow the non-monotonous boundary curve in CW direction until: 
(i) a trigger $(x',y')$ is reached, or 
(ii) the traversal falls below the \emph{floor}, on which we stop and return \emph{false}.
If a trigger is reached in State~$2$, then the goal is to rebuild a part of the monotone path by \emph{lowering} a portion of it.
First, find in the currently recorded path the last point whose height is $y'$, using a na\"ive scan from the end of the list.
We call this point $(x'',y')$ the \emph{starter}.
Note that a starter is always left of its trigger (i.e. $x'' < x'$), and a trigger is always lower than its stopper, i.e. $y' < y$.
Next, scan along a horizontal ray through the freespace, from the starter towards the trigger, which is a temporary target for our search of a canonical path.
If the trigger is reached (the ray is not blocked by a boundary), replace the suffix of the recorded path (after the starter point) with the horizontal ray to the trigger, and switch back to State~$1$. 
However, if the horizontal ray hits a free space boundary at obstruction point $(u,y')$, which can only be in a non-monotonous section of the boundary (cf. Figure~\ref{fig:mon-bdry}~(a)), then push the trigger onto the stack and repeat State~$2$ from the obstruction $(u,y')$.
If at a later point in the State~$2$ traversal, the vertical line through the ceiling's trigger point is \emph{surpassed}, then pop the trigger from the stack (e.g. ceiling update at points $8$ and $11$ in Figure~\ref{fig:ex_traversal}).

\begin{figure}
\noindent
\begin{minipage}{.6\columnwidth}
\centering
    \includegraphics[width=\columnwidth]{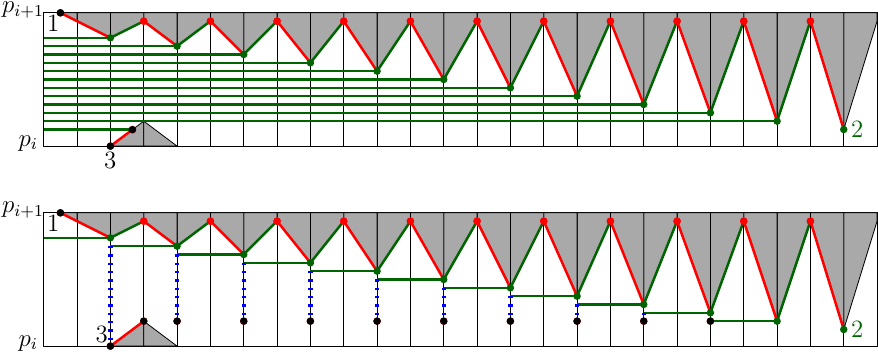} \\
	\caption{
	An example of two methods that attempt to reach trigger points, for a given row $i$ that has cascading non-freespace blockages.
	The top shows the horizontal ray shooting approach (left-to-right), that starts at point~1, then cascading trigger points to the right of 1 are reached with ray shoots, until it attempts to reach point~2 which is blocked, then it goes to point~3.
	The bottom shows the sweep approach (left-to-right) that tracks the vertical reachable space interval (blue dotted line) and starts at point~1, then cascading trigger points to the right of~1 are reached by looking at the previous cell's blue interval, until it attempts to reach point~2 which has an empty reachable space interval, so it backtracks cell-by-cell (right-to-left sweeping) to point 3 which is the first reachable point that is at the bottom of the row and blocked to the right. 
	}
	\label{fig:horiz_ray}

\end{minipage} \hfill
\begin{minipage}{.35\textwidth}
    \centering
    \hfill\includegraphics[width=\columnwidth]{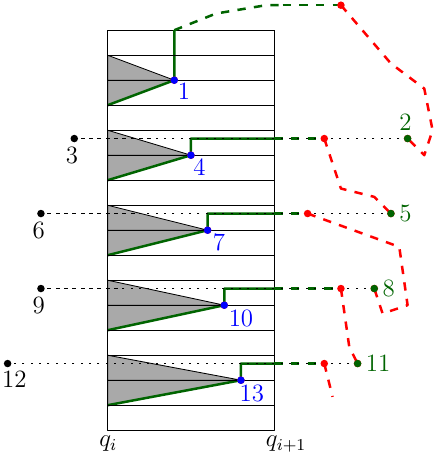} \\
	\caption{Vertical ray shooting example with cascading emission points in given column $i$.
	The canonical path reaches emission point 1.
	A vertical ray shoot from 1 reaches the top of the freespace diagram and the algorithm continues until trigger point 2 is reached.
	A horizontal ray shoot hits 3, and then the traversal reaches emission point 4.
	A vertical ray shoot from 4 hits the ceiling between 2 and 3, then goes right along the ceiling.
	The pattern continues for points 5, 6, 7, then 8, 9, 10, and finally 11, 12, 13.
	The ceilings guarantee that at most a linear amount of vertical ray shooting work can occur for a given column.
	}
	\label{fig:vert_ray}
\end{minipage}

\end{figure}

We now discuss basic properties of the greedy decider.
Cascading trigger points on the stack have monotonous coordinates, both in $x$ and in $y$.
The ceiling mechanism ensures that the search does not revisit the same stopper point.
When a trigger is reached in State~$2$, all freespace enclosed by a certain region is \emph{pruned} and will not be traversed again, i.e. the region enclosed by the monotonous path from starter to the stopper, the non-monotonous boundary from the stopper to the trigger, and above the horizontal ceiling line between the trigger and starter. 
See Figure~\ref{fig:ex_traversal} for a Greedy Decider example that shows pruning and retrieves the canonical path after only two State~$1$ traversals.

Note the following key invariants of above's algorithm that any sequence of break points, encountered in the freespace diagram, has.
\begin{enumerate}
    \item[(L)] If starter $(x',y)$ follows trigger $(x,y)$, then $x' < x$.  \label{invariant:leftness}
    \item[(B)] If trigger $(x',y')$ follows stopper $(x,y)$, then $y' < y $.\label{invariant:belowness}
\end{enumerate}

\subsubsection{Horizontal Sweeping} \label{ssec:new-sweeping}
The horizontal ray shooting in State~$2$, as described above, may perform in a single row up to a quadratic amount of work (see Figure~\ref{fig:horiz_ray} (top) for an example).
We now describe a modification that allows us to obtain an improved worst-case bound for this part.
To this end, we replace the horizontal ray shooting with the following sweep method:
Scanning the horizontal section to the right of a starter $(x,y)$ keeps records of vertical intervals 
$[y',y]$, of the form $\ceil{y-1} < y' \leq y \leq \ceil{y}$, for every passing of the free interface between two adjacent cells.
Such an interval $[y',y]$ serves the information to quickly decide if the horizontal section, from the cell with the starter point $(x,y)$, can be lowered to, say $y''$, on this cell interface. 
That is if and only if $y'' \in [y',y]$.
Hence, when a trigger point is reached, it suffices to sweep from the rightmost vertical interval in that row towards the trigger point (left-to-right), to obtain the subsequent vertical intervals.
If the trigger point falls below the range of the rightmost vertical interval, backtrack along the current canonical path suffix to find the rightmost reachable point whose scan is blocked, and continue in State~$2$.
See Figure~\ref{fig:horiz_ray} (bottom) for an example of the sweep method.

\subsubsection{Greedy Decider Analysis} \label{ssec:new-analysis}
The greedy decider may only visit a \emph{small portion} of the reachable cells in practice.
We now show that its worst-case time matches the $\O(nm)$ bound of the well known dynamic program~\cite{alt95}, whilst also using linear space.

The proposed Greedy Decider uses $\O(m+n)$ space since only the trace of the current monotone path is stored in the doubly connected list and trigger points on the stack have monotonous decreasing coordinates.
There are three types of greedy decider operations that one must analyze in order to determine an asymptotic bound: 
(i) traversal of the boundary curve (State $1$ or $2$),
(ii) vertical ray shooting (State $1$), and 
(iii) horizontal ray shooting or sweeping (State $2$).

Any point on the boundary curve is traversed at most once, due to trigger point ceilings which prevents the traversal from re-entering pruned area of the freespace (cf. point $4$ in Figure~\ref{fig:ex_traversal}).
Thus, at most $\O(mn)$ freespace boundary segments are traversed in the worst-case.

For the vertical ray shooting cost, we show a $\O(n)$ bound for the total work done in any given column.
Consider the encountered emission points in column $i$ of the freespace diagram.
Any given emission point is traversed at most once, since such points are on the boundary curve of the freespace.
Thus, it suffices to show that a cell in the column is traversed at most once by a vertical ray from unobstructed emission points beneath it.
For a given cell, consider those points beneath sorted by descending $y$-coordinates (e.g. Figure~\ref{fig:vert_ray}).
The sequence has monotonous $x$-coordinates and the emission points must be encountered in the descending $y$-order, since a canonical path that starts the $j$-th emission is above a canonical path that starts at the $(j+1)$-th emission.
We show that there is at least one active $y$-ceiling in the $y$-interval between two consecutive emission points, starting at the topmost emission point pair, $e_1$ and $e_2$.
Since $e_2$ is lower than $e_1$, it can only be reached after $e_1$ if a (stopper and) trigger are encountered.
Specifically, only if a horizontal shooting is triggered in a column right of $e_2$ and obstructed left of $e_2$.
There are three potential cases for the height of this trigger point in regard to the $y$-interval of $e_1$ and $e_2$, but $e_2$ can only be reached in one of them.
That is, if the trigger's height is beneath $e_2$, then $e_2$ is not reached since it resides in the pruned area above the ceiling.
If the trigger's height is above $e_1$, then the horizontal scan starts form a suffix after $e_1$ and cannot be obstructed in a column left of $e_2$.
Thus, $e_2$ can only be reached if the active ceiling has a height inside the $y$-interval of $e_1$ and $e_2$.
Note that the same argument applies for the $y$-interval between the $j$-th and $(j+1)$-th emission point.
Hence, the total work of vertical ray shooting in column $i$ is at most $\O(n)$ and consequently $\O(nm)$ over the entire diagram.

\begin{figure}[]
    \centering
    \includegraphics[width=\columnwidth]{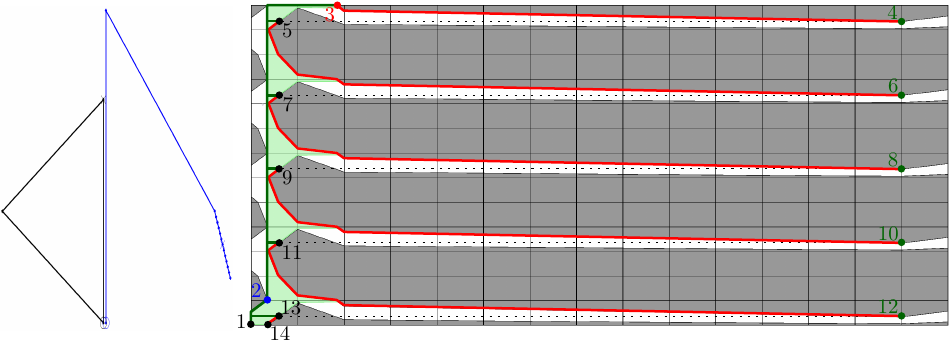}
    \caption{Example of $\O(n)$ reachable space and $\O(mn)$ Greedy Decider state $2$ work.
    {\normalfont
    The left side shows two trajectories $|P|=14$ (black, counter-clockwise triangle motion) and $|Q|=16$ (blue, starts at bottom, goes up, and then down to the right).
    The right side shows a freespace diagram for $P$ and $Q$ with reachable space in light shaded green.
    The Greedy Decider starts at $1$, goes to emission $2$, reaches stopper $3$, then cascades to triggers $4,6,8,10$ and $12$, then hits the bottom $14$ and returns $false$.
    }
    }
    \label{fig:worst_eg}
\end{figure} 

We have discussed two different methods for searching horizontal sections, and either can be used with the Greedy Decider: horizontal ray shooting or horizontal sweeping.
Clearly, every individual horizontal search takes $\O(m)$ time.

Though shooting seems natural, in the worst-case, one may encounter in a single row several cascading trigger points, each slightly lower than the previous, which results in a total work that is quadratic in $m$ for this row.
Since each cell in the given row can be traversed $\O(m)$ times, we have $\O(m^2n)$ work for horizontal shooting in total (see Figure~\ref{fig:horiz_ray} top).

Using horizontal sweeping, however, 
the canonical path stores the vertical reachable space interval for each cell on the path.
When a new trigger point is encountered, it suffices to go back to the \emph{last} stored vertical reachable space interval in that row, and then sweep left-to-right.
The last stored vertical reachable space interval in that row acts as shortcut since it contains reachable space information gathered up to that cell, i.e., we do not have to search anything to the \emph{left} of it.
Thus the vertical reachable space interval is computed at most \emph{once} for every cell interface, and at most $\O(m)$ times for a given row.
In the case of an empty vertical reachable space interval (see Figure~\ref{fig:horiz_ray} bottom), backtracking traverses every cell in the row at most once.
Hence, there is $\O(mn)$ sweeping work in total for the freespace diagram.

Putting all bounds together, the runtime of the Greedy Decider with sweeping is $\O(mn)$.
See Figure~\ref{fig:worst_eg} for a worst-case example.

\subsubsection{Greedy Decider Extension for Sub-Trajectories} \label{ssec:new-cf-gd}
We now discuss changes required to decide if $Q$ is within a distance $\tau$ of any sub-trajectory $P'$ within $P$.
Similar to the Baseline~$1$ algorithm, the goal is to search for a monotone path from anywhere on the left side to anywhere on the right side of the freespace diagram.
The target is set to any point on the right side ($x=m$).
Compute the (maximal) freespace intervals on the left side ($x=1$).
Then, for each freespace interval $[\dot{y},\ddot{y}]$,
set the \emph{floor} to $\dot{y}$, start point to $(1,\dot{y})$, and begin in State~1.
If any of the $[\dot{y},\ddot{y}]$ intervals gives a canonical path, return $true$, else return $false$.
Moreover, when testing the next lower interval at $x=1$ we use the monotonous sequence of previously encountered points on floor levels to prevent the current search from re-traversing pruned freespace,
i.e. those points are used as ceilings on the stack.
We denote the sub-trajectory greedy decider as $DP_{GD}(P,Q,\tau)$, and it runs in $\O(nm)$ time and $\O(n+m)$ space.

Similar to Baseline Algorithm $1$, we check to ensure paths are vertex aligned, and plug $DP_{GD}$ into an exact sub-trajectory nearest-neighbor algorithm, denoted $CF_{GD}(P,Q)$. 
The distance and minimal inclusion sub-trajectory are returned for the case where $DP_{GD}(P,Q,\tau) = true$ and $\tau$ is minimal.

The algorithm may finish very quickly on simple freespace instances, especially given that $m \ll n$ in our setting.
Our experiments show that, on real and synthetic data sets, the number of cell visits rarely approaches $n\cdot m$, and that typically only $\O(m)$ cells are visited per $DP_{GD}$ call.

\subsection{Hierarchical Simplification Trees (HSTs)} \label{ssec:hst}
Our proposed HST structure facilitates search space pruning during query execution by enabling the query algorithms (in Sections~\ref{ssec:algo2} and~\ref{ssec:algo3}) to quickly construct sub-trajectory candidates, from coarser to finer trajectory simplifications, based on pre-computed simplifications.
Let $\rho(i,j) = reach(\langle p_i,\ldots, p_j\rangle)$ denote the reach of the sub-trajectory between vertex $p_i$ and $p_j$ of $P$.
Recall that the greedy ball simplification gives that
$$CF(P,s(P,\tau)) \leq \tau \; \; \text{and} \; \; s(P, \rho(1,n-1))=\langle p_1, p_n \rangle.$$

We call integer $l$ a \emph{resolution} by associating $l$ to the ball radius $r(l)=2^l$ that is used for the simplification algorithm.
E.g., one may think of the set of trajectories $\{ s(P,r(l)):l \in \mathbb{Z}\}$ as various resolutions of the original trajectory $P$.
For a sub-trajectory, we are interested in the smallest integer whose simplification coincides with the line segment (spanning start and end point).

\begin{figure}
    \begin{minipage}{.48\columnwidth}
        \centering
        \includegraphics[width=\columnwidth]{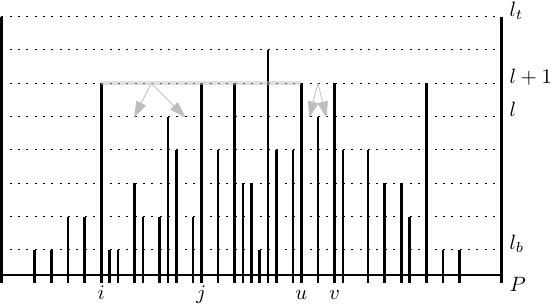}
        \caption{Example of the HST over $P$. {\normalfont Simplification $S(c)$, shown in gray, is associated to a candidate $c=([i,j],[u,v])$.  Both HST nodes $[i,j]$ and $[u,v]$ are active on resolution $l+1$, but not on $l$. All sub-trajectories $P'=\langle p_s,\ldots, p_t \rangle$ with $s \in [i,j]$ and $t \in [u,v]$ have distance $CF(S(c),P')\leq r(l+1)$. } }
        \label{fig:hst-example}
    \end{minipage}
    \hfill
    \begin{minipage}{.48\columnwidth}
        \centering
        \includegraphics[width=.7\columnwidth]{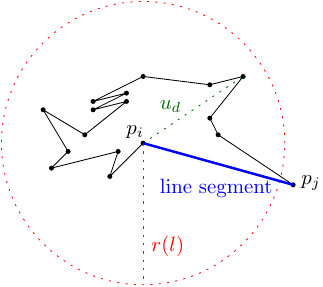}
        \caption{An example HST node at resolution level $l$, showing an unsimplified sub-trajectory $\langle p_i,\ldots, p_j \rangle$, and the corresponding simplified line segment using $s(P,r(l))$. 
        {\normalfont The dotted-line $u_d$ denotes the Euclidean distance from $p_i$ to the furthest vertex within $r(l)$.
        }
        }
        \label{fig:hst_segment_eg}
    \end{minipage}
\end{figure}

Nodes in the HST store an interval $[i,j]$, their resolution $l=\ceil{\log_2 \rho(i,j-1)}$,
and a list of children that has either zero or at least two entries.
The intervals of the children form a partition of the interval of the parent node and leaves have intervals that contain exactly two consecutive vertices of $P$ and $l=-\infty$.
For an internal node $v$, let $l(v)$ denote its resolution and $l^\dagger(v)$ the maximum resolution of its children.
We call $v$ \emph{active} on resolution $l$ if  $l^\dagger(v) < l \leq l(v)$.
See Figure~\ref{fig:hst-example} for an example of the structure.

The HST is constructed using the following top-down approach.
For the root, store the interval $[1,n]$ and $l = \ceil{ \log_2 {\rho(1,n-1)} }$.
Then recursively refine nodes with $l \neq -\infty$ by running the simplification algorithm with $r(l-1)$ on the node's sub-trajectory, i.e. $s(\langle p_i,\ldots,p_j\rangle, r(l-1))$.
For each of the obtained line segments, create one child node that contains the respective interval and compute the resolution of each child.
We denote with $l_t$ the largest and with $l_b$ the smallest resolution of internal nodes in the HST.
Clearly the HST has size $\O(n)$ and depth $D \leq \min \left\{ |P|, \O \left( \log \Delta(P) \right) \right \} $, where the spread $\Delta(P)$ denotes the ratio between the largest and smallest Euclidean distance of the set of points of $P$. 
Note that consecutive vertices have $\lVert p_i - p_{i+1}\rVert>0$.

Building any HST level, based on the previous level, takes $\mathcal{O}(n)$ time for the simplifications and reach computations.
Hence the construction time is $\mathcal{O}(n D)$.
See experiments on the obtained node degree and depth in Section~\ref{ssec:exp_results}.

\subsection{Algorithm 2 - Finding Trajectory Clusters} \label{ssec:algo2}
Algorithm~$2$ performs a nearest sub-trajectory search in the HST, and is partially based on
the Cover Tree nearest-neighbor search technique~\cite{beygelzimer2006cover}, which performs a breadth-first search and prunes branches with the help of clusters and the triangle inequality.\footnote{Using the Cover Tree on the set of $\binom{n}{2}$ sub-trajectories, similarly to Baseline~$3$, is not practical due to its large size and construction time.}
The key difference between ours and the Cover Tree algorithm is that for a given HST resolution $l$, our search uses a method, $getC(l,\C)$, that deduces a set of sub-trajectory candidates on-the-fly based on the candidates $\C$ from the previous resolution $l+1$.

Let $I(l)$ denote the set of intervals of the HST nodes active in resolution $l$.
A \emph{candidate} of this resolution is a pair of intervals 
$$c=([i,j],[u,v]) \quad \text{from} \; \; I(l) \times I(l), \; \; \text{with} \; \; i < j, \; \; u < v, \; \; \text{and} \; \; i \le u.$$
We call $[i,j]$ the start interval, $[u,v]$ the end interval, and its associated simplification $S(c)$ starts at $p_i$ and ends at $p_u$, and we have $CF(S(c),\langle p_i, \ldots ,p_u\rangle) \leq r(l) $.
The associated cluster $C(c)$ of $c$ consists of the sub-trajectories $P'=\langle p_s, \ldots, p_t \rangle$, with $s \in [i,j]$ and $t \in [u,v]$, each of which has distance $CF(S(c),P') \leq r(l)$ (see Figure~\ref{fig:hst-example}).

Procedure $getC(l,\C)$ generates, from a candidate set $\C$ at resolution $l+1$, a new candidate set at resolution $l$ by replacing those intervals of candidates that are active on $l+1$ but not on $l$ with one candidate pair for each (newly active) child node.

The basic idea of the nearest sub-trajectory search is as follows.
Loop over the resolutions of the HST, from $l_t$ down to the leaf level.
At each resolution $l$ the following is done.
Sub-trajectory candidates are constructed for the resolution based on the remaining candidates from the previous resolution.
For each candidate $c \in \C$, compute $CF(S(c),Q)$ and set $\alpha$ to be the smallest of these distances.
Then any $c \in \C$ with $CF(S(c),Q)> \alpha + 2r(l)$ is discarded from the set of candidates.
Once all resolutions have been searched ($l=-\infty$), $\C$ contains all sub-trajectories of $P$ that realize the same (nearest-neighbor) distance to $Q$, e.g. we report the inclusion minimal result if $|\C| > 1$.

\setcounter{algocf}{1}

{\blue
\begin{algorithm}
    \SetAlgoLined
    \KwResult{sub-trajectory distance and $P'$ }
    \KwData{$\C=\{([1,n],[1,n])\}$}
    \For{$l \leftarrow l_t-1$ \textbf{down} \KwTo $l_b$}{
        $\C \longleftarrow getC(l,\C)$\;
        $\alpha \longleftarrow$ smallest $CF(S(c),Q)$ with $c \in \C$\;
        discard $c \in \C$ \textbf{if} $\alpha + 2r(l) < CF(S(c),Q)$\;
    }
    Return $\alpha$, $P' \in \C$
\caption{HST Search $CF_{HST}(Q)$}
\label{alg:a3}
\end{algorithm}
}

\begin{figure*}\centering
	\includegraphics[width=.98\textwidth,height=6.0cm]{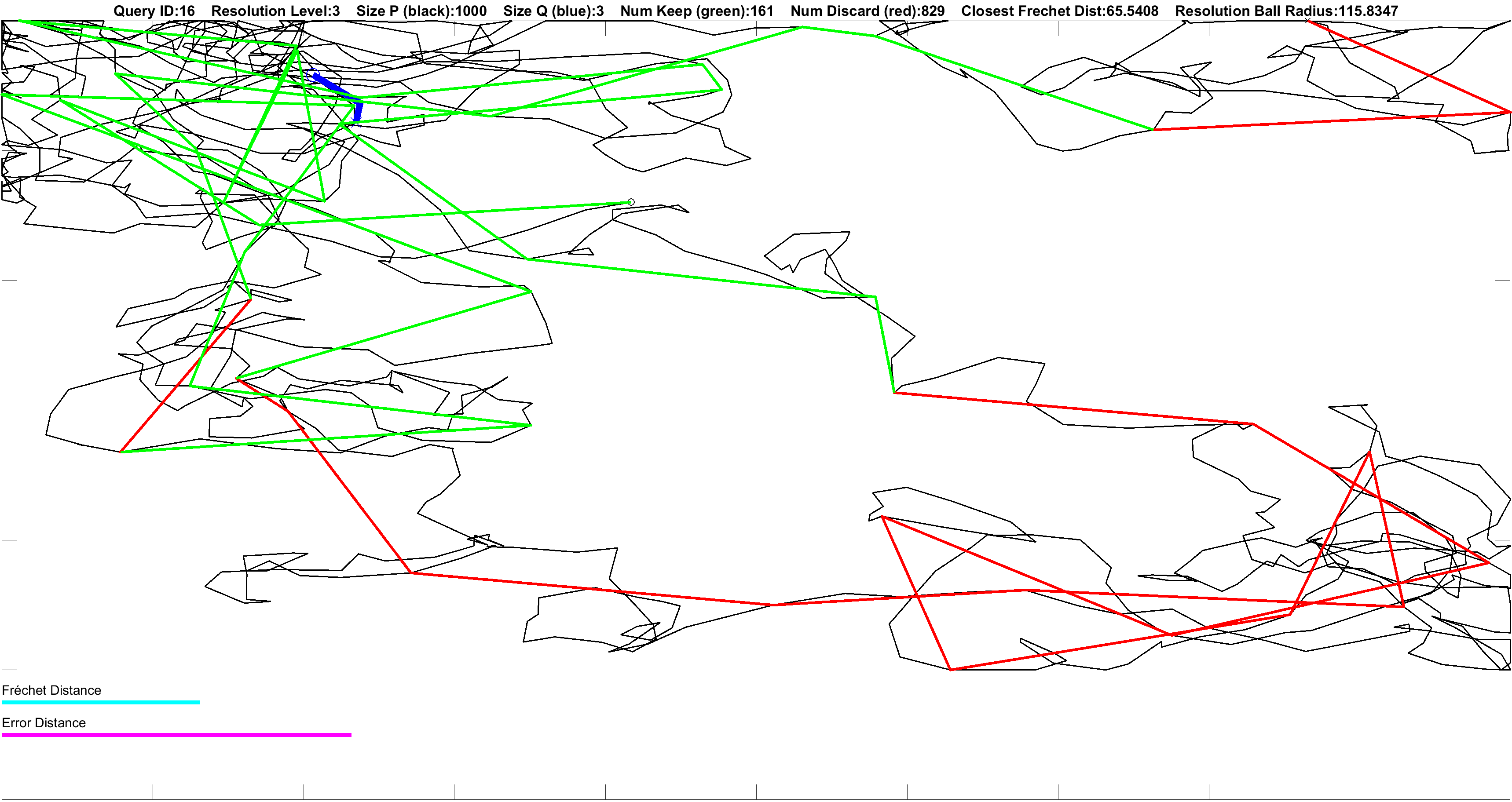} \\
	\vspace{0.1cm}
	\includegraphics[width=.98\textwidth,height=6.0cm]{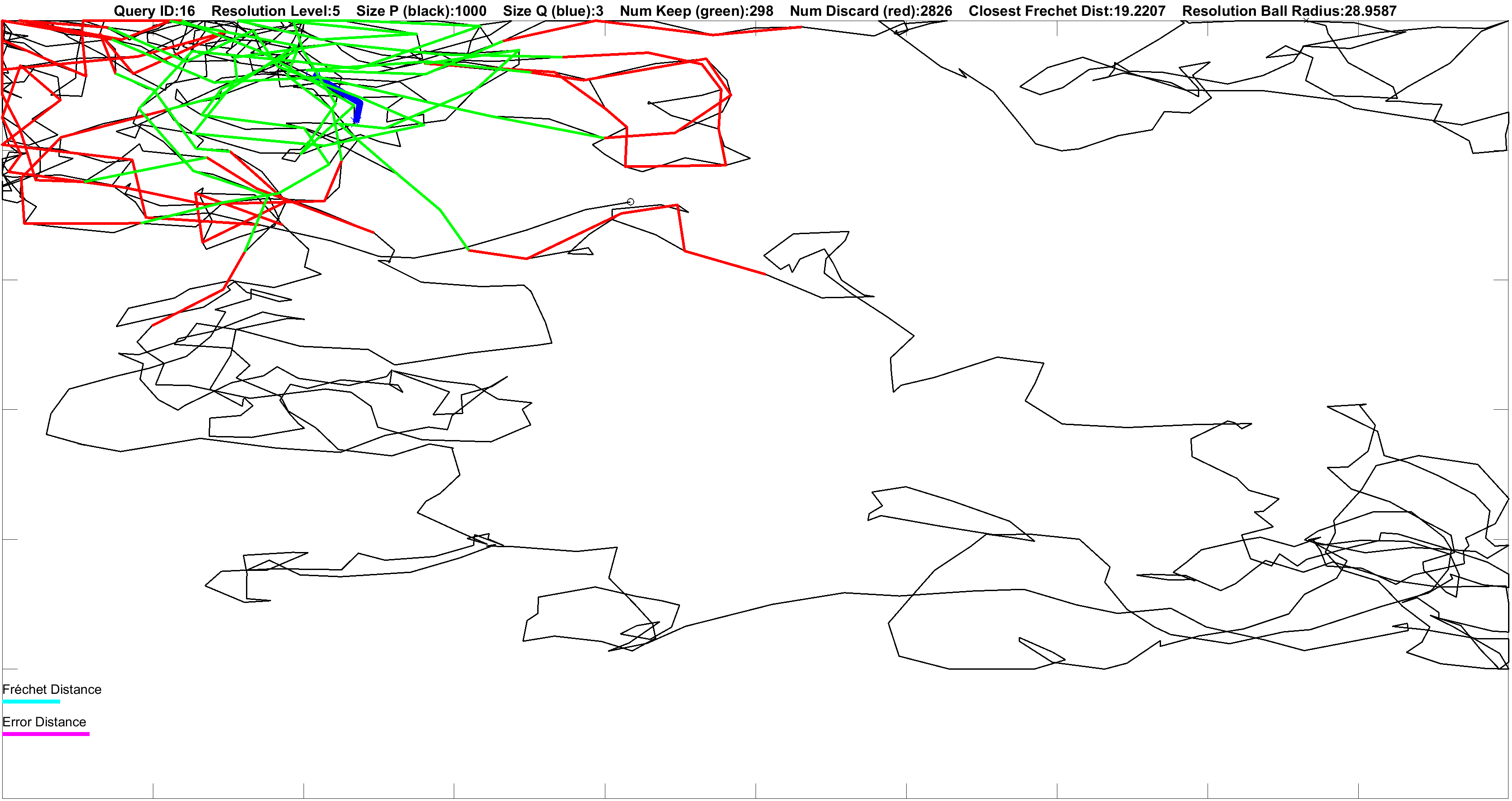} \\
	\vspace{0.1cm}
	\includegraphics[width=.98\textwidth,height=6.0cm]{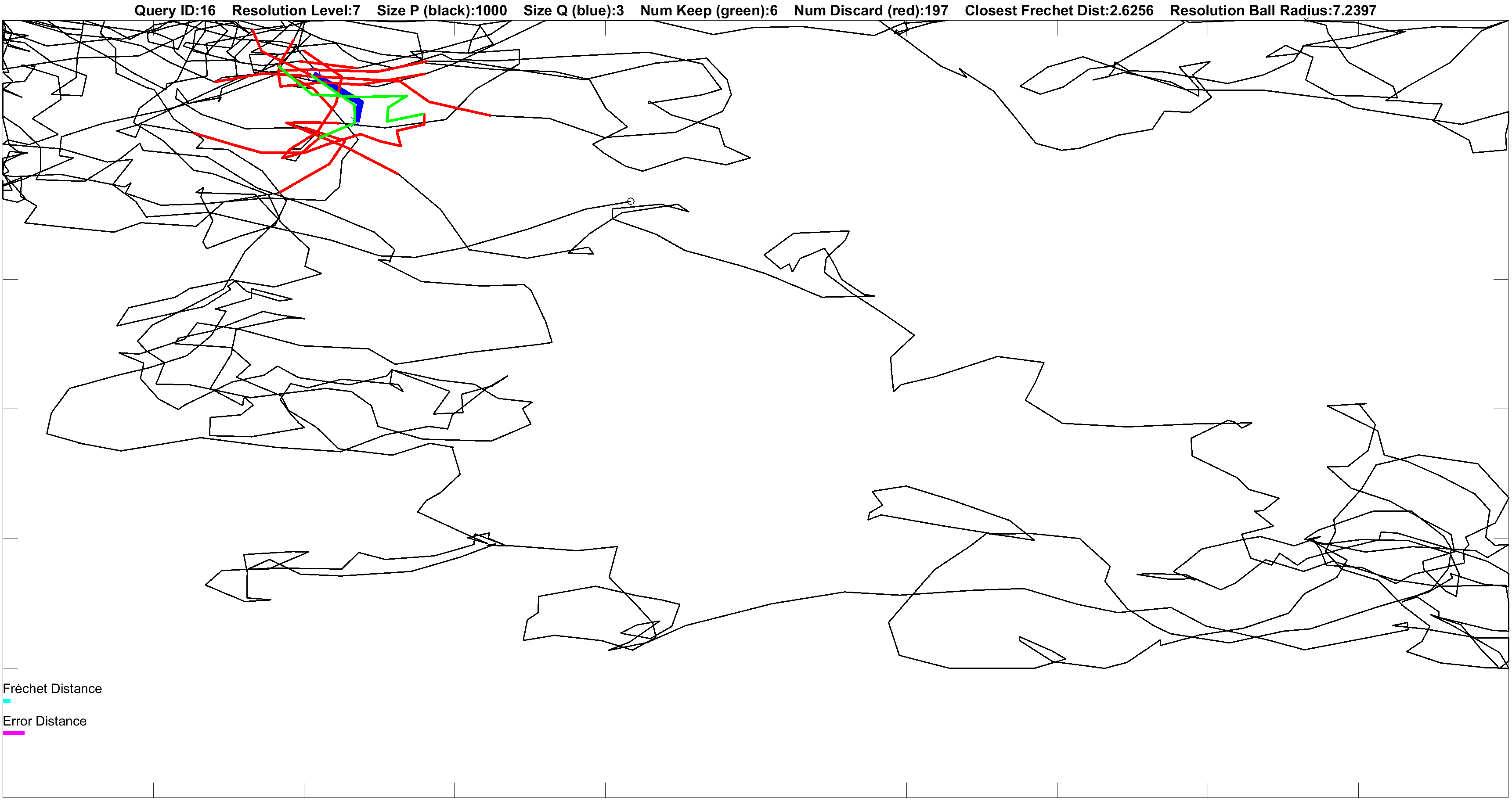} 
	\caption{Algorithm 2 query example on the Synthetic (high) $|P|=10^3$ data set (see Section~\ref{ssec:exp_setup}).
	The plots show the input trajectory $P$ (black), query trajectory $Q$ (blue), and candidate trajectories that are pruned (red) and kept (green) as the search goes from larger (top) to smaller (bottom) resolution $l$ errors.
	}
	\label{fig:algo2_eg}
\end{figure*}


Figure~\ref{fig:algo2_eg} shows an example query for Algorithm~$2$.
On larger resolution $l$, the candidates have longer segments and are further away from the non-simplified vertices of $P$ that they `cover'.
As the $l$ ball radius is reduced, candidate segments reduce in length and more closely cover their respective $P$ vertices.
At each resolution $l$, candidates that are far enough away from $Q$ are pruned.

\subsubsection{Algorithm 2 Analysis}
The HST search is efficient if candidate pruning is effective and candidate sizes $|S(c)|$ remain small, since this reduces the time to compute the CF distances.
For example, processing any of the $|getC(l,\C)|$ candidates of resolution $l$ takes time 
$$\O( \lambda_l m \log(\lambda_l m)), \; \; \text{where} \; \; \lambda_l =\max \{|S(c)|:c \in getC(l,\C)\}$$ 
is the largest size of an (encountered) simplification on resolution $l$.
Candidates $S(c)$ tend to stay small in size at upper HST levels since they have a larger simplification error and hence contain fewer vertices.
Also, if the nearest neighbor $P'$ has small size, then even $S(c)$ at lower HST levels tend to remain relatively small in size.
(E.g. the bottom of Figure~\ref{fig:query_results} in Section~\ref{sec:exp} shows this behavior).
However, Algorithm~$2$ has the same drawback as the Baseline~$3$ CCT algorithm.
If there are many (exact) nearest sub-trajectories to test, then pruning is less effective and more $CF$ calls are issued (cf. Section~\ref{ssec:exp_results}).

In the worst-case, no candidates are pruned and the HST is very unbalanced, meaning most levels have close to $n$ nodes.
There are at most a quadratic number of pairwise candidates for each level $l$ where $CF(P,Q)$ is computed, which results in $\mathcal{O}((l_t - l_b) \cdot n^2)$ distance calls.
Fortunately, experiments show that this algorithm performs much fewer distance calls, especially when $|P|$ is smaller and has lower intrinsic dimensionality.

\subsection{Algorithm 3 - Putting All Together} \label{ssec:algo3}
The improved HST search algorithm overcomes issues mentioned in Algorithm~$2$ by combining ideas from Algorithm~$1$ and $2$, introducing a switch that changes the search technique if $|\C|$ gets too large, and incorporating other heuristics that reduce computations in practice. 

Algorithm~$3$ 
contains two parts.
Part~$1$ (lines $1$-$7$) is a modification of Algorithm~$2$ that uses faster upper and lower bounds on the continuous Fréchet distance to more quickly search candidates on \emph{parent} resolutions, and stops if $|\C|$ gets too large.
Part~$2$ (lines $8$-$17$) searches any remaining \emph{parent} resolution and the \emph{leaf} level by calling Algorithm~$1$, but on (typically) small candidate $c$ sizes at various simplification resolutions.

\begin{algorithm}
    \SetAlgoLined
    \KwResult{sub-trajectory distance and $P'$ }
    \KwData{$l \leftarrow l_t;~\C=\{([1,n],[1,n])\}$}
    \While( \tcp*[h]{\textbf{Part~$1$}}){$l > l_b $}{ 
        $l \leftarrow l-1$\;
        $\C \leftarrow getC(l,\C)$\;
        \textbf{if} $l = l_b$ \textbf{or} $|\C| > M$ \textbf{then} \textbf{break}\;
        $\alpha \longleftarrow$ smallest $UB_1(S(c),Q)$ with $c \in \C$\;
        discard $c \in \C$ \textbf{if} $\alpha + 2r(l) < \max(LB_1(S(c),Q),LB_2(S(c),Q))$\;
    }
    $C \longleftarrow aggregateC(\C)$  \tcp*[h]{\textbf{Part~$2$}}\;
    \For{$l' \leftarrow l$ \textbf{down} \KwTo $l_b$}{
        $\alpha$, $P' \longleftarrow$ smallest $CF_{GD}(S(c),Q)$ with $c \in C$\;
        \If{$l' > l_b$}{
            $\overline{C} \leftarrow \emptyset$\;
            append inclusion maximal paths to $\overline{C}$ by calling $DP_{GD}(S(c),Q,\alpha + r(l'))$ with $c \in C$\;
            $C \longleftarrow aggregateC(\overline{C})$\;
        }
    }
    Return $\alpha$, $P'$
\caption{NearestSubTrajectory$(Q)$}
\end{algorithm}

\subsubsection{Part 1 - Pruning using Heuristics}
We use upper and lower bound computations for $CF$ that run, on-the-fly, in linear time or, using additional space, even in $\O(1)$ and $\O(\log n)$ time.
Line $5$ in Part~$1$ sets $\alpha$ to the smallest upper bound $UB_1(S(c),Q)$ from~\cite{bringmann2019walking}, which essentially uses the lengths of trajectory $S(c)$ and $Q$, from their respective midpoints, and the Euclidean distance between the midpoints on $S(c)$ and $Q$, to arrive at an upper bound (see Figure $7$ in~\cite{bringmann2019walking}). 
Note that a linear time pre-processing step allows one to store the cumulative lengths of the line segments.
Part~$1$ (line $6$) then discards $c$ if $\alpha + 2r(l)$ is less than any of two lower bound computations.
The first lower bound $LB_1(S(c),Q)$ is a constant time bound, also from~\cite{bringmann2019walking} which uses the same information as the upper bound computation.
However, $LB_1$ is a very weak bound (e.g., returns $0$) if the trajectory $S(c)$ is long.
To improve the chance of discarding $c$, we also call a stronger lower bound, $LB_2(S(c),Q)$, from~\cite{gudmundsson2021practical}, which computes the two start/end point Euclidean distances and bounding boxes of $S(c)$ and $Q$ in $\mathcal{O}(\log n)$ time using an augmented search tree.

Part~$1$ (line $4$) is the circuit-breaker that checks if it is worthwhile or not to continue checking processing parent levels using upper/lower bounds.
If the candidate set $|\C|$ becomes larger than a threshold $M$ (e.g., a constant), then Part~$1$ stops early.

There are two additional heuristic modifications that reduce the number of candidates during the search.
%
%
The first modification is how candidates are generated by the $getC$ function (line~$3$) for the next lower resolution $l$.
We take advantage of a useful HST property, namely that a given candidate $c$ generated at resolution $l$ can be \emph{pushed down} and described at an even lower resolution ($l-1,l-2$, etc.), if the nodes that describe $S(c)$ are the \emph{same} from one resolution to the next.
For example, in Figure~\ref{fig:hst-example}, $c$ cannot be pushed down further since the nodes that describe it change from one resolution to the next.
However, when analyzing real and synthetic inputs, we observed that it is often the case that a given candidate $c$ has the same set of nodes describing it for multiple resolutions.
So, when generating a set of candidates for resolution $l$, some of the candidates may be tagged with even lower resolutions, but we only process (i.e., compute upper/lower bound distances) for those candidates at level $l$.
This gives the algorithm an opportunity to further reduce the smallest upper bound $\alpha$ before processing those candidates that were pushed down, and hence may promote discarding those candidates.

The second modification reduces the error resolution $r(l)$ when checking if a given candidate $c$ can be discarded (line~$6$).
Recall that a given HST node $u$ at resolution $l$ represents a line segment interval obtained from the trajectory simplification algorithm $s(P,\tau)$, and that $u$'s error is \emph{at most} $r(l)$.
At HST construction time, for each node $u$, we store the Euclidean distance $u_d$ from the start vertex to the furthest vertex that is within the $r(l)$ ball, i.e. $u_d \le r(l)$.
See Figure~\ref{fig:hst_segment_eg} for an example.
For a given candidate $c$ at level $l$, we compute the maximum $u_d$ from the candidate's nodes, and use this value instead of $r(l)$ when performing the pruning check.
Analysis of real data sets shows that $u_d$ is often much less than $r(l)$, and hence this promotes discarding candidates.

\subsubsection{Analysis of Part 1}
If the query on $P$ is fortunate, meaning the upper/lower bound pruning is effective and the circuit-breaker is avoided, then all resolutions are processed in Part~$1$ (beside the leaf level which is always processed in Part~$2$).
Part~$1$ is also particularly favorable if $|S(c)|$ is small. 
Using pre-computed data for the bound computations on every HST level, the worst case time of Part~$1$ entails, on each HST level, at most $\O(M)$ bound computations.
Thus, using error bound $r(l)$ for pruning, Part~$1$ can be executed in $\mathcal{O}(  (l_t-l_b)\cdot M \cdot \log n)$ time.

\subsubsection{Part 2 - Batching Candidates}
We now discuss Part~$2$ (lines $8$-$17$), which abandons the metric pruning of sub-trajectory clusters inherent in Part~$1$, and instead opts for a freespace pruning method that calls the proposed Algorithm~$1$ greedy decider with candidates that have decreasing resolutions.

The first step (line $8$) aggregates existing pairwise candidates into a smaller set that is more fortunate for doing searches in the freespace diagram.
In $aggregateC(\C)$, the union of candidate intervals is computed, which combines intervals that overlap.
The new aggregated candidate set still covers all original intervals, but the aggregation can result in a (much) smaller set of candidates.
For example, on a given level, the result of $aggregateC$ on $\{ ([1,3],[1,3]),([3,4],[7,8]),([5,6],[9,10])\}$ yields the candidate $c=([1,3],[9,10])$ and $S(c)=\langle p_1,p_3,p_4,\ldots,p_{10} \rangle$.
(Note that for Part~$2$, the simplified trajectory $S(c)$ generated for a given $c=([i,j],[u,v])$ starts at vertex $i$ and ends at vertex $v$, unlike during Part~$1$.)
This `batching process' results in fewer calls to the greedy decider, and avoids unnecessary work since overlapping candidate intervals are eliminated.

The second step (lines $9-16$) loops from the HST resolution $l$, that Part~$1$ stopped at, down to the leaf level and does the following.
First, in line $10$, for each candidate $c \in C$, compute $CF_{GD}(S(c),Q)$ and set $\alpha$ and the sub-trajectory result $P'$ to the smallest $CF_{GD}(S(c),Q)$, where $P'$ is the minimal inclusion result.
If we are at the leaf level, then the continuous Fréchet distance $\alpha$ and nearest sub-trajectory $P'$ are returned.
Otherwise, $\alpha$ plus the resolution error is used as an upper bound to assist in computing a candidate set for the next finer resolution (line $13$), which is then aggregated (line $14$).
We describe candidate set generation in detail below, as well as additional heuristics to speed-up the algorithm.

There are different ways to compute a candidate set for the next finer resolution (line $13$).
For example, one could again simply discard a candidate $c \in C$, if $\alpha + 2r(l') < CF_{GD}(S(c),Q)$.
However, remaining candidates retain their start $[i,j]$ and end $[u,v]$ intervals, even if those parts become further from $Q$ as the resolution decreases.

Instead, we opt for a candidate set generation method (line $13$) that \emph{reduces} remaining candidate interval lengths, and prunes away candidates that are too far.
This improved pruning method calls $DP_{GD}(S(c),Q,\alpha + r(l'))$ for each aggregated $c \in C$, which returns \emph{maximal inclusion} paths, i.e. the new candidates for the next lower resolution level.
There are three possible outcomes when calling $DP_{GD}$: 
(i)   $c$ is pruned since there is no longer a canonical path in the free space, 
(ii)  $S(c)$ is \emph{reduced} in length since the maximal inclusion path that is returned is shorter, or 
(iii) $c$ is retained and the maximal inclusion path is not reduced in length.
Since the closest sub-trajectory can be no further than $\alpha + r(l')$, the call on a candidate with the closest result will always return a maximal inclusion path and thus remain in the set.
As the resolution is reduced in Part~$2$, the distance $\alpha + r(l')$ becomes smaller, hence improving the pruning effectiveness when calling $DP_{GD}$.
It is important to note that maximal inclusion paths (rather than minimal inclusion) are returned from $DP_{GD}$, to ensure that only non-reachable space that is \emph{further} than $\alpha + r(l')$ is pruned.

Three additional heuristics are applied to the $CF_{GD}(S(c),Q)$ calls in line~$10$ that estimate $\alpha$:
\begin{itemize}
    \item Uniformly randomly choose a small number ($\approx\ceil{\log |C|}$) of candidates in $C$ to call $CF_{GD}(S(c),Q)$ and determine $\alpha$.
    \item Limit decision calls in $CF_{GD}(S(c),Q)$, using the current resolution $l'$, to avoid unnecessary precise estimation of $\alpha$.
    \item If $LB_2(S(c),Q) > \alpha$ or $DP_{GD}(S(c),Q,\alpha) = false$, then do not check $CF_{GD}(S(c),Q)$.
\end{itemize}
The first two heuristics typically perform fewer computations nearer the HST root and more computations nearer the HST leaf level, and can result in a larger $\alpha$, thus not violating pruning correctness.

\subsubsection{Analysis of Part 2}

The aggregated candidates in $C$ have a total size of no more than $|P|$.
So, the runtime of Part~$2$ is $\O((l_t - l_b) \cdot mn \log mn)$, based on calling $CF_{GD}$ up to $(l_t - l_b)$ times.
It uses $\mathcal{O}(n+m)$ space since  $\O( |C|+\max_{c \in C}|S(c)| )$ additional memory is used.

\subsubsection{Modification for Approximate Nearest-Neighbor}
Note that Algorithm~$3$ is easily adjusted (lines $5$ and $10$) to allow for results with additive $\varepsilon^+$ or relative $(1+\varepsilon^*)$ errors.
Pass in $\varepsilon^+$ or $\varepsilon^*$ as part of the query, and if $\alpha - r(l) > 0$, then do the following check.
For additive errors, if $(\alpha + r(l)) - (\alpha - r(l)) \le \varepsilon^+$, then stop and return the $c$ and $\alpha$.
For relative errors, if $(\alpha + r(l)) / (\alpha - r(l)) \le 1+\varepsilon^*$, then stop and return the $c$ and $\alpha$.

\subsubsection{Algorithm 3 Analysis and Pruning Effectiveness}

It is difficult to choose a satisfactory circuit breaker mechanism to switch from Part~$1$ to Part~$2$, since Part~$2$ can perform much less work than its worst-case runtime on realistic inputs.
I.e., the precise tradeoff point between Part~$1$ and Part~$2$ is elusive due to heuristics and the grouping of candidates.
We set the switch to $M = 1000$ (line $4$) for our query experiments (Section ~\ref{sec:exp}), which was determined based on the following experimental evaluation.
Nearest query experiments were run on the $|P|=10^4$ Pigeon and Football real data sets for $M = 500,1000,1500,2000,2500,3000,3500, $ and $4000$. $M = 1000$ gave the best runtime for both data sets.
This switch value seems beneficial since it gives Part~$1$ an opportunity to discard candidates before switching over to Part~$2$.

Algorithm~$3$ reduces performance issues associated with the Baseline algorithms.
Baseline $1$ 
needs to check all reachable cells, whereas Algorithm~$3$ uses linear space and heuristics which can greatly reduce freespace cell checks.
Baseline $2$ performs many loops checking irrelevant parts of $P$, but Algorithm~$3$ prunes parts of $P$ that are too far.
Baseline $3$ has a quadratic size in the sub-trajectory setting and an issue with multiple exact closest results that require additional distance computations, whereas Algorithm~$3$ uses the linear size HST, and avoids the multiple result problem by using candidate aggregation.

Algorithm~$3$ also reduces deficiencies of Algorithms~$1$~\&~$2$.
Algorithm~$1$'s heuristics can greatly reduce cell-checks, however, it still has to search irrelevant parts of unsimplified $P$, but the pruning of Algorithm~$3$ at higher levels of simplification discards unnecessary parts fast and early.
Algorithm~$3$ performs fewer computations than Algorithm~$2$ 
since its Part~$1$ is faster than Algorithm~$2$.

\section{Experiments} \label{sec:exp}
We experimentally evaluate the scalability, effectiveness, and efficiency of our proposed algorithms and HST data structure versus three Baseline algorithms.
We measure HST construction runtimes and statistics, and for queries we measure runtimes and candidate sizes against both real and synthetic $2$D data sets.
All experiment code and data sets are publicly available.\footnote{See \url{https://github.com/japfeifer/frechet-queries} for code and data sets.}

The experiment setup is described first, followed by our results.
\subsection{Experiment Setup} \label{ssec:exp_setup}
Experiments are performed on a laptop computer with an Intel Core i$7$-$10875$H CPU 
and $64$GB RAM, using a \emph{single threaded} Matlab implementation (interpreted programming language) on a Windows~$10$ $64$-bit OS.
Experiments use a precision binary search algorithm (similar to~\cite{bringmann2019walking}) for the continuous Fréchet distance (and sub-trajectory version).


\paragraph{Real Data Sets.}
Three real data sets are used for experimentation.
The first data set tracks Homing Pigeons~\cite{pigeon16}, from release sites to a home site, and contains $131$ trajectories each having an average of $970.0$ vertices.
The second data set tracks European Football players on the pitch~\cite{soccer15}, with trajectories representing the movement of a player when they have possession of the ball, and has $18$,$034$ trajectories, each with $203.4$ vertices on average.
The third data set contains $180$,$736$ Taxi cab trajectories~\cite{taxiA11,taxiB10} gathered by GPS as they traverse the streets of Beijing, China.

Input trajectories $P$ are constructed for $|P|=500$, $10^3$, $5 \times 10^3$, $10^4$, $5 \times 10^4$, and $10^5$ for each real data set. 
The assembly of $P$ works as follows.
A uniformly randomly chosen trajectory is removed from the data set and appended to initially empty $P$.
Then, remove the next trajectory in the data set whose start vertex is the closest Euclidean distance to the current end vertex of $P$, and append it to the end of $P$.
This process continues until the desired $|P|$ is achieved.

The real data sets are difficult to search since they contain many clusters of overlapping sub-trajectories that are similar.
The Pigeon data set has bird flight paths that have many similarities since the birds all share a common home site.
The Football data set is slightly more challenging since its trajectories are confined to a small area, and players often use the football pitch in similar ways.
Notably, the Taxi data set is the most difficult to search, since its trajectories are in a relatively small area and taxi routes are often on similar roads, hence many large-size disjoint sub-trajectories often belong to big clusters that are very close to each other.
We conduct additional experiments on the Taxi data set with even larger $|P|=$ $5 \times 10^5$ and $10^6$ input trajectories, which is our hardest test, since there are up to one million vertices containing large contiguous disjoint sub-trajectories that are similar.

\begin{figure}[h]
\hspace{-0.7cm}

\centering
\begin{tikzpicture}

\begin{groupplot}[
     group style = {group size = 4 by 4,
                    horizontal sep=0.15cm,
                    vertical sep=0.1cm,},
     height=3.3cm,
     width=3.8cm,
     grid style=dashed,
     stack plots = false,
     xmode = log,
     xtick=data,
     xticklabels={,,},
     xmin=250, xmax=200000,
     ymode = log,
     ymin=40, ymax=1500,
     ytick={10,100,1000},
     ymajorgrids=true,
     log ticks with fixed point,
     log origin = infty,
    ]

\nextgroupplot[
xtick style={draw=none},
yticklabels={,$10^{6}$,$10^{7}$,$10^{8}$},
ymin=100000, ymax=50000000,
ytick={100000,1000000,10000000,100000000},
ylabel={Runtime (ms)},
y label style={at={(axis description cs:-.31,0)},anchor=south},
axis y discontinuity=parallel,
axis x line=top, 
legend style={at={(2.16,1.42)},anchor=north},
legend columns = 2,
]

\addplot[color=red,mark=otimes,]
coordinates {(500,2982000)(1000,23834000)};

\addplot[color=blue,mark=square,]
coordinates {(500,39)(1000,62)(5000,162)(10000,233)(50000,1360)(100000,2601)};

\legend{Baseline 3}
\addlegendentry{HST}

\nextgroupplot[
xtick style={draw=none},
yticklabels={,,},
ymin=100000, ymax=50000000,
ytick={100000,1000000,10000000,100000000},
axis y discontinuity=parallel,
axis x line=top, 
]

\addplot[color=red,mark=otimes,]
coordinates {(500,1437000)(1000,11531000)};

\addplot[color=blue,mark=square,]
coordinates {(500,13)(1000,27)(5000,111)(10000,268)(50000,1221)(100000,2314)};

\nextgroupplot[
xtick style={draw=none},
yticklabels={,,},
ymin=100000, ymax=50000000,
ytick={100000,1000000,10000000,100000000},
axis y discontinuity=parallel,
axis x line=top, 
]

\addplot[color=red,mark=otimes,]
coordinates {(500,851000)(1000,5812000)};

\addplot[color=blue,mark=square,]
coordinates {(500,15)(1000,32)(5000,146)(10000,282)(50000,1631)(100000,3091)};

\nextgroupplot[
xtick style={draw=none},
yticklabels={,,},
ymin=100000, ymax=50000000,
ytick={100000,1000000,10000000,100000000},
axis y discontinuity=parallel,
axis x line=top, 
]

\addplot[color=red,mark=otimes,]
coordinates {(500,2628000)(1000,22015000)};

\addplot[color=blue,mark=square,]
coordinates {(500,11)(1000,21)(5000,131)(10000,275)(50000,1362)(100000,2580)};


\nextgroupplot[
xtick style={draw=none},
yticklabels={,$10^{1}$,$10^{2}$,$10^{3}$},
ymin=4, ymax=5000,
ytick={1,10,100,1000,10000},
axis x line=bottom,
]

\addplot[color=red,mark=otimes,]
coordinates {(500,2982000)(1000,23834000)};

\addplot[color=blue,mark=square,]
coordinates {(500,39)(1000,62)(5000,162)(10000,233)(50000,1360)(100000,2601)};

\nextgroupplot[
xtick style={draw=none},
yticklabels={,,},
ymin=4, ymax=5000,
ytick={1,10,100,1000,10000},
axis x line=bottom,
]

\addplot[color=red,mark=otimes,]
coordinates {(500,1437000)(1000,11531000)};

\addplot[color=blue,mark=square,]
coordinates {(500,13)(1000,27)(5000,111)(10000,268)(50000,1221)(100000,2314)};

\nextgroupplot[
xtick style={draw=none},
yticklabels={,,},
ymin=4, ymax=5000,
ytick={1,10,100,1000,10000},
axis x line=bottom,
]

\addplot[color=red,mark=otimes,]
coordinates {(500,851000)(1000,5812000)};

\addplot[color=blue,mark=square,]
coordinates {(500,15)(1000,32)(5000,146)(10000,282)(50000,1631)(100000,3091)};

\nextgroupplot[
xtick style={draw=none},
yticklabels={,,},
ymin=4, ymax=5000,
ytick={1,10,100,1000,10000},
axis x line=bottom,
]
	
\addplot[color=red,mark=otimes,]
coordinates {(500,2628000)(1000,22015000)};

\addplot[color=blue,mark=square,]
coordinates {(500,11)(1000,21)(5000,131)(10000,275)(50000,1362)(100000,2580)};


\nextgroupplot[
xtick style={draw=none},
group style = {vertical sep=0.15cm,},
ymin=7, ymax=28,
ytick={10,15,20,25},
ymode = normal,
ylabel={Depth $D$},
ylabel shift = 0.05cm,
]

\addplot[color=blue,mark=square,]
coordinates {(500,21)(1000,22)(5000,22)(10000,22)(50000,22)(100000,22)};

\nextgroupplot[
xtick style={draw=none},
group style = {vertical sep=0.15cm,},
yticklabels={,,},
ymin=7, ymax=28,
ytick={10,15,20,25},
ymode = normal,
]

\addplot[color=blue,mark=square,]
coordinates {(500,12)(1000,14)(5000,15)(10000,15)(50000,15)(100000,15)};

\nextgroupplot[
xtick style={draw=none},
group style = {vertical sep=0.15cm,},
yticklabels={,,},
ymin=7, ymax=28,
ytick={10,15,20,25},
ymode = normal,
]

\addplot[color=blue,mark=square,]
coordinates {(500,15)(1000,17)(5000,19)(10000,20)(50000,24)(100000,24)};

\nextgroupplot[
xtick style={draw=none},
group style = {vertical sep=0.15cm,},
yticklabels={,,},
ymin=7, ymax=28,
ytick={10,15,20,25},
ymode = normal,
]

\addplot[color=blue,mark=square,]
coordinates {(500,10)(1000,10)(5000,13)(10000,14)(50000,14)(100000,14)};


\nextgroupplot[
group style = {vertical sep=0.15cm,},
xlabel={Pigeon},
xtick={100,1000,10000,100000},
xticklabels={,$10^{3}$,$10^{4}$,$10^{5}$},
yticklabels={,$10^{1}$,$10^{2}$,$10^{3}$,},
ymin=3, ymax=2000,
ytick={1,10,100,1000,10000},
ylabel={Max Degree},
ylabel shift = -0.05cm,
]

\addplot[color=blue,mark=square,]
coordinates {(500,14)(1000,15)(5000,18)(10000,26)(50000,45)(100000,79)};

\nextgroupplot[
group style = {vertical sep=0.15cm,},
xlabel={Football},
xtick={100,1000,10000,100000},
xticklabels={,$10^{3}$,$10^{4}$,$10^{5}$},
yticklabels={,,},
ymin=3, ymax=2000,
ytick={1,10,100,1000,10000},
]

\addplot[color=blue,mark=square,]
coordinates {(500,7)(1000,10)(5000,13)(10000,17)(50000,28)(100000,66)};

\nextgroupplot[
group style = {vertical sep=0.15cm,},
xlabel={Synth. low},
xtick={100,1000,10000,100000},
xticklabels={,$10^{3}$,$10^{4}$,$10^{5}$},
yticklabels={,,},
ymin=3, ymax=2000,
ytick={1,10,100,1000,10000},
]

\addplot[color=blue,mark=square,]
coordinates {(500,6)(1000,5)(5000,6)(10000,8)(50000,7)(100000,8)};

\nextgroupplot[
group style = {vertical sep=0.15cm,},
xlabel={Synth. high},
xtick={100,1000,10000,100000},
xticklabels={,$10^{3}$,$10^{4}$,$10^{5}$},
yticklabels={,,},
ymin=3, ymax=2000,
ytick={1,10,100,1000,10000},
]

\addplot[color=blue,mark=square,]
coordinates {(500,9)(1000,12)(5000,52)(10000,104)(50000,466)(100000,884)};

\end{groupplot}

\end{tikzpicture}
\caption{Construction runtimes for Baseline~$3$ CCTs and HST data structures (top), HST depth (middle), and HST maximum degree (bottom).
{\normalfont
Columns (from left to right) show the Pigeon and Football real data sets and Synthetic low and high intrinsic dimensionality data sets, with $|P|=$ $500$, $10^3$, $5 \times 10^3$, $10^4$, $5 \times 10^4$, and $10^5$ for each (x-axis is log scale).
Baseline~$3$ experiments are run up to $|P|=10^3$ (quadratic-size CCT construction runtime issues occur for $|P|>10^3$).
HST mean degree is in $[2.3,2.5]$ for all data sets.
}
}
\label{fig:hst_results}
\end{figure}
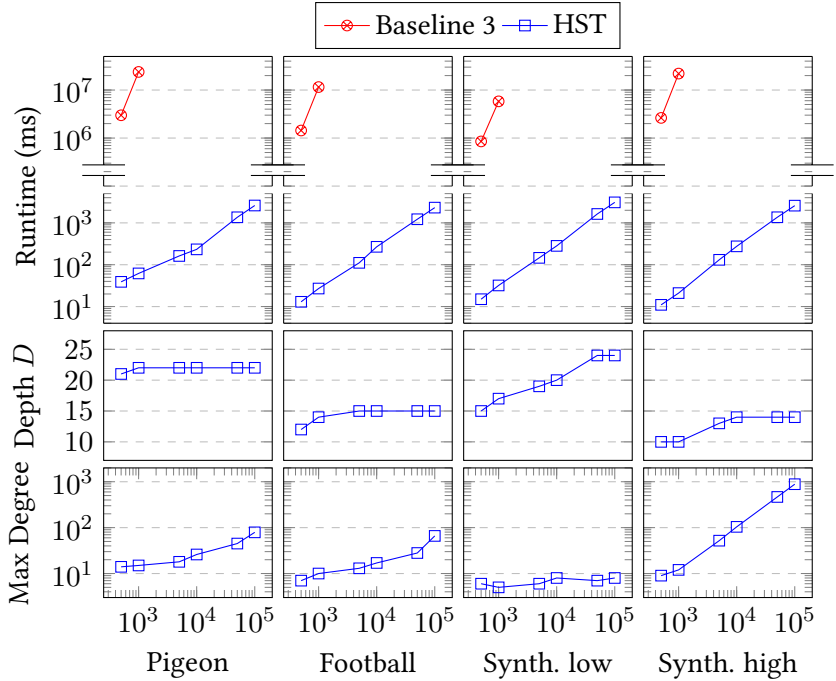

\paragraph{Synthetic Data Sets}
Two types of synthetic input trajectories $P$ are generated for experimentation.
The first type contains an input curve with \emph{low} intrinsic dimensionality, i.e., it has longer segments and is straighter, resulting in sub-trajectories that have less overlap.
The second type contains an input curve with \emph{high} intrinsic dimensionality (more overlap), and has more pronounced directional changes and is also confined to a small area.
We use the low and high setting of the tool from~\cite{gudmundsson2021practical} to generate synthetic input trajectories of size  $|P|=$ $500$, $10^3$, $5 \times 10^3$, $10^4$, $5 \times 10^4$, and $10^5$.

\paragraph{Query Generation}
Queries are generated as follows for a given input trajectory $P$.
Locate a contiguous sub-trajectory on the input trajectory that is between $2$ and $2 \log n$ contiguous vertices in size (uniform random size and location in $P$).
Copy the sub-trajectory, perturb the vertices up to $3\%$ of its $reach$, then uniformly randomly translate it up to $5\%$ of its $reach$, which results in a query trajectory.
Repeat the process until $1$,$000$ queries are generated.

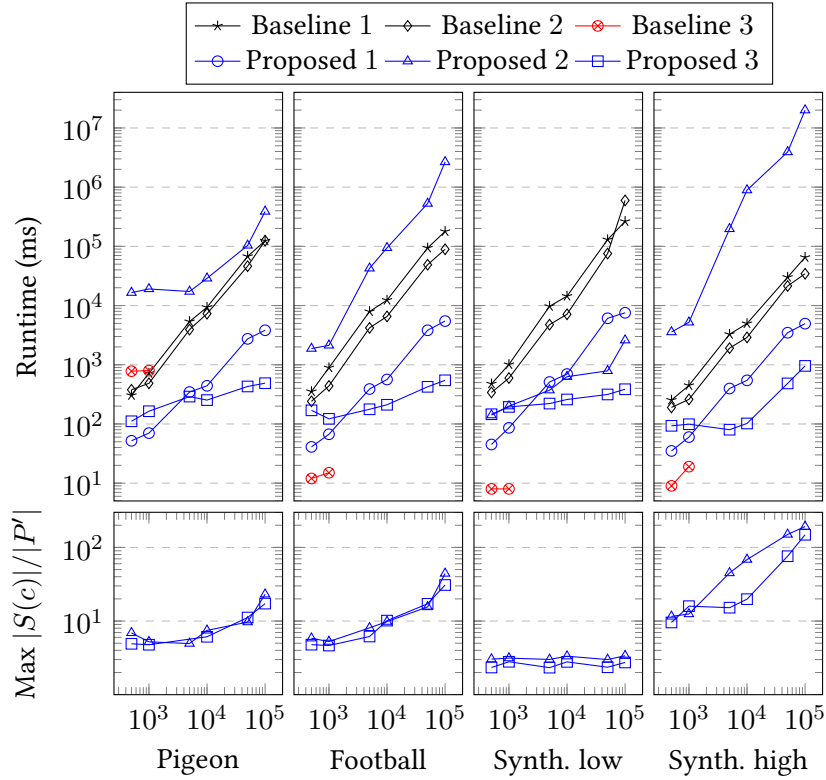
\begin{figure}[ht]

\centering

\begin{tikzpicture}

\begin{groupplot}[
     group style = {group size = 4 by 2,
                    horizontal sep=0.15cm,
                    vertical sep=0.15cm,},
     height=7.0cm,
     width=3.8cm,
     grid style=dashed,
     stack plots = false,
     xmode = log,
     xtick=data,
     xticklabels={,,},
     xmin=250, xmax=200000,
     ymode = log,
     ymin=1, ymax=300,
     ytick={1,10,100,1000},
     ymajorgrids=true,
     log ticks with fixed point,
     log origin = infty,
    ]


\nextgroupplot[
xtick style={draw=none},
yticklabels={,$10^{1}$,$10^{2}$,$10^{3}$,$10^{4}$,$10^{5}$,$10^{6}$,$10^{7}$},
ymin=5, ymax=40000000,
ytick={1,10,100,1000,10000,100000,1000000,10000000,100000000},
ylabel={Runtime (ms)},
]

\addplot[color=black,mark=star,]
coordinates {(500,304)(1000,718)(5000,5370)(10000,9350)(50000,67745)(100000,124193)};

\addplot[color=black,mark=diamond,]
coordinates {(500,378)(1000,486)(5000,3927)(10000,7291)(50000,46229)(100000,124873)};

\addplot[color=red,mark=otimes,]
coordinates {(500,779)(1000,798)};

\addplot[color=blue,mark=o,]
coordinates {(500,52)(1000,70)(5000,346)(10000,441)(50000,2738)(100000,3825)};

\addplot[color=blue,mark=triangle,]
coordinates {(500,16525)(1000,19073)(5000,17272)(10000,28852)(50000,103718)(100000,388514)};

\addplot[color=blue,mark=square,]
coordinates {(500,111)(1000,164)(5000,290)(10000,253)(50000,430)(100000,488)};

\nextgroupplot[
xtick style={draw=none},
yticklabels={,,},
ymin=5, ymax=40000000,
ytick={1,10,100,1000,10000,100000,1000000,10000000,100000000},
legend style={at={(1.1,1.22)},anchor=north},
legend columns = 3,
]

\addplot[color=black,mark=star,]
coordinates {(500,353)(1000,891)(5000,7929)(10000,12342)(50000,93947)(100000,178810)};

\addplot[color=black,mark=diamond,]
coordinates {(500,241)(1000,439)(5000,4191)(10000,6554)(50000,48962)(100000,89352)};

\addplot[color=red,mark=otimes,]
coordinates {(500,12)(1000,15)};

\addplot[color=blue,mark=o,]
coordinates {(500,41)(1000,67)(5000,389)(10000,564)(50000,3825)(100000,5468)};

\addplot[color=blue,mark=triangle,]
coordinates {(500,1878)(1000,2127)(5000,42592)(10000,94364)(50000,529206)(100000,2665701)};

\addplot[color=blue,mark=square,]
coordinates {(500,170)(1000,122)(5000,177)(10000,211)(50000,424)(100000,546)};

\legend{Baseline $1$}
\addlegendentry{Baseline $2$}
\addlegendentry{Baseline $3$}
\addlegendentry{Proposed $1$}
\addlegendentry{Proposed $2$}
\addlegendentry{Proposed $3$}

\nextgroupplot[
xtick style={draw=none},
yticklabels={,,},
ymin=5, ymax=40000000,
ytick={1,10,100,1000,10000,100000,1000000,10000000,100000000},
]

\addplot[color=black,mark=star,]
coordinates {(500,476)(1000,1011)(5000,9707)(10000,14417)(50000,129795)(100000,263982)};

\addplot[color=black,mark=diamond,]
coordinates {(500,340)(1000,596)(5000,4760)(10000,7125)(50000,75553)(100000,596924)};

\addplot[color=red,mark=otimes,]
coordinates {(500,8)(1000,8)};

\addplot[color=blue,mark=o,]
coordinates {(500,45)(1000,86)(5000,511)(10000,697)(50000,6098)(100000,7550)};

\addplot[color=blue,mark=triangle,]
coordinates {(500,141)(1000,200)(5000,372)(10000,628)(50000,790)(100000,2597)};

\addplot[color=blue,mark=square,]
coordinates {(500,146)(1000,194)(5000,221)(10000,259)(50000,315)(100000,386)};

\nextgroupplot[
xtick style={draw=none},
yticklabels={,,},
ymin=5, ymax=40000000,
ytick={1,10,100,1000,10000,100000,1000000,10000000,100000000},
]

\addplot[color=black,mark=star,]
coordinates {(500,253)(1000,448)(5000,3270)(10000,4971)(50000,30118)(100000,65304)};

\addplot[color=black,mark=diamond,]
coordinates {(500,190)(1000,260)(5000,1906)(10000,2901)(50000,21393)(100000,34463)};

\addplot[color=red,mark=otimes,]
coordinates {(500,9)(1000,19)};

\addplot[color=blue,mark=o,]
coordinates {(500,35)(1000,60)(5000,397)(10000,545)(50000,3479)(100000,4947)};

\addplot[color=blue,mark=triangle,]
coordinates {(500,3573)(1000,5213)(5000,197280)(10000,891191)(50000,3923099)(100000,19971655)};

\addplot[color=blue,mark=square,]
coordinates {(500,93)(1000,99)(5000,80)(10000,102)(50000,484)(100000,956)};


\nextgroupplot[
xlabel={Pigeon},
xtick={100,1000,10000,100000},
xticklabels={,$10^{3}$,$10^{4}$,$10^{5}$},
yticklabels={,$10^{1}$,$10^{2}$},
ylabel={Max $|S(c)|/|P'|$},
height=4.0cm,
]

\addplot[color=blue,mark=triangle,]
coordinates {(500,6.91)(1000,5.22)(5000,4.95)(10000,7.47)(50000,9.69)(100000,23.11)};

\addplot[color=blue,mark=square,]
coordinates {(500,4.91)(1000,4.77)(4.96)(10000,6.12)(50000,11.17)(100000,17.26)};

\nextgroupplot[
xlabel={Football},
xtick={100,1000,10000,100000},
xticklabels={,$10^{3}$,$10^{4}$,$10^{5}$},
yticklabels={,,},
height=4.0cm,
]

\addplot[color=blue,mark=triangle,]
coordinates {(500,5.84)(1000,5.26)(5000,8.08)(10000,9.83)(50000,15.54)(100000,44.26)};

\addplot[color=blue,mark=square,]
coordinates {(500,4.78)(1000,4.65)(5000,6.17)(10000,10.14)(50000,17.15)(100000,30.87)};

\nextgroupplot[
xlabel={Synth. low},
xtick={100,1000,10000,100000},
xticklabels={,$10^{3}$,$10^{4}$,$10^{5}$},
yticklabels={,,},
height=4.0cm,
]

\addplot[color=blue,mark=triangle,]
coordinates {(500,3.04)(1000,3.14)(5000,3.01)(10000,3.33)(50000,3.00)(100000,3.38)};

\addplot[color=blue,mark=square,]
coordinates {(500,2.34)(1000,2.80)(5000,2.33)(10000,2.79)(50000,2.36)(100000,2.74)};

\nextgroupplot[
xlabel={Synth. high},
xtick={100,1000,10000,100000},
xticklabels={,$10^{3}$,$10^{4}$,$10^{5}$},
yticklabels={,,},
height=4.0cm,
]

\addplot[color=blue,mark=triangle,]
coordinates {(500,11.57)(1000,12.47)(5000,44.98)(10000,68.41)(50000,150)(100000,192)};

\addplot[color=blue,mark=square,]
coordinates {(500,9.57)(1000,15.93)(5000,15.19)(10000,19.82)(50000,75.98)(100000,148.62)};

\end{groupplot}
\end{tikzpicture}
\caption{Nearest-neighbor sub-trajectory query results, showing averages over $1$,$000$ queries, for baseline and proposed algorithms.
{\normalfont
Rows denote query runtime in ms (top), and maximum candidate $|S(c)|$ size as a factor of the sub-trajectory result $|P'|$ size (bottom).
Columns (from left to right) show the Pigeon and Football real data sets and Synthetic low and high intrinsic dimensionality data sets, with $|P|=$ $500$, $10^3$, $5 \times 10^3$, $10^4$, $5 \times 10^4$, and $10^5$ for each (x-axis is log scale).
Baseline $2$ is a $2$-approximation.
Some runtimes for slower methods show averages over $100$ (or less) queries since they take too long to execute (i.e. $> 20$ seconds per query).
Baseline~$3$ experiments are run up to $|P|=$ $10^3$ (quadratic-size CCT construction runtime issues occur for $|P|>10^3$).
}
}
\label{fig:query_results}
\end{figure}

\subsection{Experiment Results} \label{ssec:exp_results}
A comparison of construction time for the HST and the CCT data structure is shown in Figure~\ref{fig:hst_results}.
HSTs for $|P|=10^5$ take three seconds to construct.
CCT construction experiments on $|P|=5 \times 10^3$ or larger cannot be run due to the quadratic size issue and unreasonable construction time.
Compared to CCTs, HST construction is on average more than $10^5$ times faster.
HST depths $D$ are a small factor of $\log n$, and although the maximum degree grows with $|P|$, 
the average degree is in the interval $[2.3,2.5]$ on all data sets.

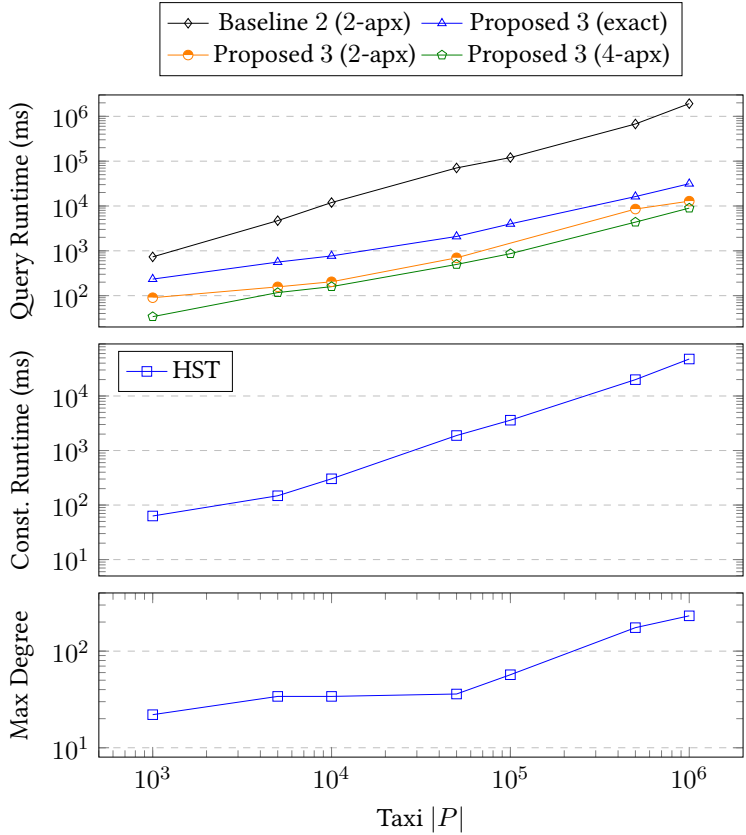
\begin{figure}[ht]

\centering

\begin{tikzpicture}[scale=.9]

\begin{groupplot}[
     group style = {group size = 1 by 3,
                    horizontal sep=0.25cm,
                    vertical sep=0.25cm,},
     height=5.0cm,
     width=11.0cm,
     grid style=dashed,
     stack plots = false,
     xmode = log,
     xtick=data,
     xticklabels={,,},
     xmin=500, xmax=2000000,
     ymode = log,
     ymin=1, ymax=300,
     ytick={1,10,100,1000},
     ymajorgrids=true,
     log ticks with fixed point,
     log origin = infty,
    ]


\nextgroupplot[
xtick style={draw=none},
yticklabels={,$10^{2}$,$10^{3}$,$10^{4}$,$10^{5}$,$10^{6}$},
ymin=20, ymax=3000000,
ytick={10,100,1000,10000,100000,1000000,10000000},
ylabel={Query Runtime (ms)},
legend style={at={(0.5,1.4)},anchor=north},
legend columns = 2,
]

\addplot[color=black,mark=diamond,]
coordinates {(1000,733)(5000,4712)(10000,11931)(50000,70606)(100000,120397)(500000,674831)(1000000,1920877)};

\addplot[color=blue,mark=triangle,]
coordinates {(1000,234)(5000,561)(10000,770)(50000,2090)(100000,3993)(500000,16220)(1000000,31399)};

\addplot[color=orange,mark=halfcircle*,]
coordinates {(1000,90)(5000,157)(10000,204)(50000,698)(1500)(500000,8542)(1000000,12823)};

\addplot[color={green!50!black},mark=pentagon,]
coordinates {(1000,34)(5000,117)(10000,159)(50000,497)(100000,864)(500000,4365)(1000000,8973)};

\legend{Baseline $2$ ($2$-apx)}
\addlegendentry{Proposed $3$ (exact)}
\addlegendentry{Proposed $3$ ($2$-apx)}
\addlegendentry{Proposed $3$ ($4$-apx)}


\nextgroupplot[
legend pos=north west,
xtick style={draw=none},
yticklabels={,$10^{1}$,$10^{2}$,$10^{3}$,$10^{4}$},
ymin=5, ymax=90000,
ytick={1,10,100,1000,10000,100000},
ylabel={Const. Runtime (ms)},
]

\addplot[color=blue,mark=square,]
coordinates {(1000,63)(5000,148)(10000,303)(50000,1883)(100000,3597)(500000,19940)(1000000,47645)};

\legend{HST}


\nextgroupplot[
height=4.0cm,
xlabel={Taxi $|P|$},
xtick={100,1000,10000,100000,1000000},
xticklabels={,$10^{3}$,$10^{4}$,$10^{5}$,$10^{6}$},
yticklabels={$10^{1}$,$10^{2}$,},
ymin=8, ymax=400,
ytick={10,100,1000},
ylabel={Max Degree},
]

\addplot[color=blue,mark=square,]
coordinates {(1000,22)(5000,34)(10000,34)(50000,36)(100000,57)(500000,175)(1000000,232)};

\end{groupplot}
\end{tikzpicture}
\caption{Resuts on the Taxi~\cite{taxiA11,taxiB10} data set that compare Baseline~$2$ vs. Proposed~$3$ query times (top), HST construction times (middle), and HST maximum degree (bottom).
{\normalfont
The top shows the time of $2$-approximate queries using Baseline~$2$ and time of Proposed~$3$ for exact, $2$-approximate, and $4$-approximate queries.
Some of the Baseline~$2$ data points show averages over $100$ (or fewer) queries since execution of the full query set took too long.
Input trajectory sizes are $|P|=10^3$, $5 \times 10^3$, $10^4$, $5 \times 10^4$, $10^5$, $5 \times 10^5$, and $10^6$ (x-axis is log scale).
The HST depth $D$ is $26$ over all input trajectory sizes.
}
}
\label{fig:subtraj_taxi_results}
\end{figure}

Figure~\ref{fig:query_results} shows the nearest sub-trajectory query runtimes for the baselines and proposed methods, and the results align with our analysis in Sections~\ref{sec:base} and~\ref{sec:new}.
Our proposed algorithm~$3$ has faster query times and scales better compared to others as $|P|$ increases, 
and the amount of work performed per $P$ and $Q$ vertex goes down to a constant when the input data sets become large.
Several of the Baseline~$1$~$\&$~$2$ and Proposed~$2$ experiments had to be run with a smaller number of queries, since they were taking much longer than $20$ seconds per query to execute, whereas the Proposed~$3$ performed well under a variety of data sets and input trajectory sizes, including the most difficult synthetic data set with high intrinsic dimensionality.
Interestingly, the Proposed~$2$ performs second-best with the low intrinsic dimensionality Synthetic trajectories, but worst for the other data sets which have higher intrinsic dimensionality, which suggests it is sensitive to this measure.
The Baseline~$3$ algorithm (the CCT-based solution) performs best on Football and synthetic data sets for $P=$ $500$ and $10^3$, due to its favorable clustering of the inputs, however, the CCT data structure size is quadratic in $n$ and hence impractical for larger $|P|$.
Surprisingly, Baseline~$3$ performs worse than Proposed~$3$ for the Pigeon $P=500$ and $10^3$ data sets, since its pruning is less effective in this case (cf. Figure~\ref{fig:realistic_inputs}~(c)).

Our proposed algorithm~$3$ encounters typically only small candidate trajectories in the search, i.e., maximum candidate sizes $|S(c)|$ are typically close to the result size $|P'|$.
This shows that the Proposed~$3$ prunes $P$ well, which results in faster query times since it does computations on smaller candidates.

\subsubsection{Approximate Queries}\label{ssec:new-apx-queries}
Figure~\ref{fig:subtraj_taxi_results} shows Taxi data set HST construction and query times for Baseline~$2$ and Proposed~$3$ algorithms.
Recall that the Taxi data set is the hardest to search. 
The HST construction time and maximum degree show similar patterns as $P$ increases in size, when compared to the other real and synthetic data sets.
For our largest input trajectory, $|P| = 10^6$, the HST construction runtime is only $48$ seconds, and its depth $D$ of $26$ is close to $\log_2 |P|$.
The Proposed~$3$ query results show increasingly faster runtimes as one goes from exact, to $2$-approximate and $4$-approximate queries.
E.g., for $|P|=10^4$, the exact search is $770$ms, the $2$-apx search is $204$ms, and the $4$-apx search is $159$ms.
Though the time for exact queries increases from $|P| = 10^3$ to $|P| = 10^6$, the slope for Proposed~$3$ is smaller than the slope of Baseline~$2$.
For example, Baseline~$2$ (2-apx) is between $3.1$ ($|P|=10^3$) to $61.2$ ($|P|=10^6$) times slower than Proposed~$3$ (exact).
These experimental results show that Proposed~$3$ scales better than our baseline and other proposed algorithms.

\section{Future Work} \label{sec:future}

We are interested in improving the analysis of proposed Algorithm~$2$ (Trajectory Clusters) and bounding the runtime based on an intrinsic dimensionality measure such as the expansion constant~\cite{kargerR02}.
The good performance of this algorithm on the Synthetic-low data set suggests that the underlying intrinsic dimensionality of $P$ is an important indicator of runtime.

It is also interesting to study other heuristics that improve practical runtimes for the proposed Algorithm~$3$.

\paragraph*{Acknowledgements}
This work was supported under the Australian Research Council Discovery Projects funding scheme (project number DP180102870).

\printbibliography

\end{document}